# The role of green ammonia in meeting challenges towards a sustainable development in China

Hanxin Zhao*

*Faculty of Technology, Policy and Management, Delft University of Technology, Jaffalaan 5, 2628BX Delft, the Netherlands*

* Corresponding author, *Email: h.zhao-1@tudelft.nl*, *hzhao8163@gmail.com*

# Abstract

This paper discusses the adoption of a green ammonia economy in meeting challenges in China's sustainable development. First, key challenges in China's energy transition, industry decarbonziation and regional sustainable development are explored. The coal-dominated energy consumption has placed great obstacles in achieving energy transition and led to massive CO2 emission since the large-scale industrialization. The high dependency on oil and gas import has threatened the energy security. A DEA model is applied for obtaining green total factor productivities of China's six administrative regions, with which, imbalanced and unsustainable regional development is identified. Second, the role of green ammonia in meeting the sustainability challenges is analysed. Ammonia is examined to be a flexible and economic option for large-scale hydrogen transport and storage. Co-firing ammonia in coal power generation at 3% rate is evaluated as an option for achieving low-carbon transition by 2030. The adoption of a green ammonia economy in China is discussed from energy, environmental and economic aspects. The practice can decline fossil energy consumption, enhance energy security, and facilitate renewable energy delivery and storage, industry decarbonization, and regional development. We assume the findings and results contribute to addressing sustainability challenges and realizing a hydrogen economy in China.

# Key words

*green ammonia; hydrogen economy; ammonia economy; energy transition; China;*

# 1. Introduction

## 1.1 Green ammonia in carbon neutrality and sustainable development

China pushes towards peaking emission by 2030 and reaching neutrality by 2060 [1]. The announcement represents a concrete vision for realizing low-carbon and sustainable development. For the past 2 decades, investments in wind and solar energy grew significantly and accounts for one-third of the world's total capacity [2]. However, renewable resources are mostly distributed in Northern and Western China, while the economic clusters are generally located in Eastern and Southern China [3]. The future energy pattern of China features a distinct spatial imbalance between renewable energy supply and demand. The gird inflexibility and insufficient transmission has led to a growing divergence between installed capacity and actual power generation from renewable energy [4]. As a result, despite the rapid development in renewable energy, China is expected to remain the largest consumer of fossil energy by 2050 [5]. The challenges call for a concerted effort from green hydrogen, thanks to the technology of 'Power to Gas' (P2G). Lately, China step a move to green hydrogen with mid and long-term targets set in 2022 for achieving carbon reduction goals and supporting the energy transition [6]. However, to date, the large-scale use of green hydrogen faces challenges in shipping and storing hydrogen, in addition to the high production cost [7]. With the feature of being a clean fuel and an efficient and safe hydrogen carrier, green ammonia is increasingly paid attention in terms of playing a key role in a hydrogen economy [7]. Green ammonia is expected to solve the problem of single hydrogen energy, and therefore is regarded as 'hydrogen 2.0' [8]. The use of green hydrogen to decarbonize the traditional ammonia sector is highlighted in China's 2021-2035 Hydrogen Development Plan [6]. However, the role of green ammonia in achieving China's carbon neutrality and sustainable development has not been sufficiently recognized. An assessment of the development of green ammonia supply chains in meeting sustainability challenges in China is necessary.

## 1.2 Research on the adoption of the hydrogen and ammonia economy

An increasing number of recent studies have discussed the hydrogen economy. Many studies discussed institutional conditions for the hydrogen economy [9-11]. For example, Ashari et al. explored the industrial, political and social determinants of a hydrogen economy [9]. Harichandan et al. analysed the green hydrogen supply chain in India and proposed policy directives for developing a hydrogen economy [10]. Chu et al. proposed supply and demand plan for green hydrogen in South Korea [11]. Some studies have evaluated policy impacts for developing the hydrogen economy [12, 13]. For example, Li et al. evaluated impacts of subsidy policies of green hydrogen in China [12]. Some studies have contributed to the hydrogen market [14, 15]. For example, Jia et al. explored joint bidding strategies for electricity and hydrogen markets [14]. An increasing number of studies have contributed to discussing the feasibility of a hydrogen economy [16-20]. For example, Lee et al. explored technological and institutional vulnerabilities towards a hydrogen economy in South Korea [19]. Palacios et al. evaluated the potential need and feasibility of adopting a hydrogen economy in Mexico [20]. Khan et al. analysed the sustainable development of the hydrogen economy in the Mid East [16]. Huang et al. assessed the potential of hydrogen development in cities in China based on a supply-demand-policy model [17]. However, limited studies have discussed the adoption of a hydrogen economy in a region or country from a comprehensive perspective. Hong et al. evaluated the sustainability of a hydrogen economy in South Korea from energy, economic and environmental aspects [18], while a deep insight into the main challenges in the country's energy transition and decarbonisation was less considered in the evaluation.

With the role of green ammonia in the hydrogen economy valued, supply chains and the economy of green ammonia have received more attention. Though, a large proportion of studies up to date concern production process and operation of green ammonia [21, 22]. For example, Gujarathi et al. compared ammonia produced using solar energy and natural gas from economic and environmental aspects [21]. Wu et al. proposed a method to address electricity-hydrogen-ammonia matching problem under load fluctuations in green ammonia production [22]. With regard to the ammonia supply chain and associated economy, a great majority of studies have contributed to techno-economic assessment of the supply chain [23-26]. For example, Zhao et al. evaluated impacts of economic incentives on investment in green ammonia production in China [24]. Tu et al. explored ammonia transportation using pipeline in China [25]. Many studies on the green ammonia economy have concentrated on the feasibility of the economy [26-29]. For example, Sekhar et al. discussed the significance of green ammonia in decarbonizing the energy sector [27]. Morlanes et al. analysed the potential roles of green ammonia as an energy carrier [28]. Galimova et al. assessed the feasibility of importing green ammonia from the North Africa and South America to the Europe [29]. Besides, market structures in the green ammonia economy was explored by the work [26]. However, the role of green ammonia subject to energy, economic and environmental concerns has not been well researched. In particular, the feasibility of developing the green ammonia economy has been drawn little attention, with studies on green ammonia mainly focusing on techno-economic assessment of green ammonia production and supply chains.

## 1.3 Research goal and contribution

In summary, the development of hydrogen energy in China is expected to speed up for realizing a low-carbon and sustainable development, due to the climate change and environmental degradation. Whereas, the use of hydrogen in a large scale faces challenges due to the low density and high flammability. Green ammonia derived from green hydrogen is increasingly drawn attention as a hydrogen carrier and clean fuel in a hydrogen economy. However, the role of green ammonia in addressing China's challenges towards sustainability has not drawn sufficient attention by policymakers and recent studies in academia. This paper aims to study the feasibility of developing a green ammonia economy under sustainable development challenges in China. The main contributions are summarized as follows:

(1) Distinct with recent studies on the feasibility of adopting a green ammonia economy, this study first takes a closer and comprehensive exploration on key challenges in China's energy transition, industry decarbonisation and regional sustainable development. The findings form a solid basis for evaluating the role of green ammonia in achieving sustainability in China and also provide insights regarding the main concerns in China's sustainable development.
(2) The role of green ammonia as a hydrogen carrier and clean fuel, as well as the potential supply and demand from applicable sectors in China are analysed concerning the feasibility of developing green ammonia supply chains in China. Based on the evaluations and the key challenges identified, the adoption of a green ammonia economy in China is comprehensively discussed from energy, environmental and economic aspects. The study provides references and pathways in addressing challenges towards a sustainable society and achieving a hydrogen economy in China.

## 1.4 Structure of the paper

The remainder of this paper is arranged as follows: the framework for analysing sustainability of energy systems is introduced in Section 2. Section 3 discusses challenges in China's energy transition, industry decarbonisation and sustainable development. Section 4 discusses the role of green ammonia in meeting the challenges. Section 5 presents the conclusions and limitations of the work.

# 2. Methods

## 2.1 Framework for analysing sustainability of energy systems

Energy management has become a complex subject, as the transition to a low-carbon energy system is not only a technical issue, but much a social one [30]. With the shift of general reorganization, the focus of energy transition lies in the sustainable development of energy systems which pays attention to a clean and efficient energy supply and the support to economic development. The energy-environment-economy (3E) concept has been widely used for energy policy making and governance, as shown in Fig. 1. For example, 3E was introduced as the 'three pillars' of Japan's energy policy [31]. The framework and definition of each pillar vary from case to case. The 3E according to the international discourse falls in the area of energy supply security, energy equity and energy greening [32]. Japan considered the addition of safety, since nuclear safety has been an integral part of the energy policy [31].

As the sustainable energy development requires a comprehensive coordination between energy development, economic growth and carbon emission reduction, and by referring to 3E frameworks from international discourse and countries, the definition of 3E in this study is as follows. The dimension of energy notes securing a stable energy supply, while the dimensions of environment and economy emphasizing environmental suitability and economic efficiency. In addition, trade-offs are needed to achieve the sustainable energy development with priorities decides the goals to focus on, considering that the three dimensions cannot be achieved together. In this study, securing a stable energy supply is regarded as the priority with the purpose of satisfying massive energy demand and enabling economic development in China. Meanwhile, energy security should be aligned with the goals of improving economic efficiency and carbon emission reduction.

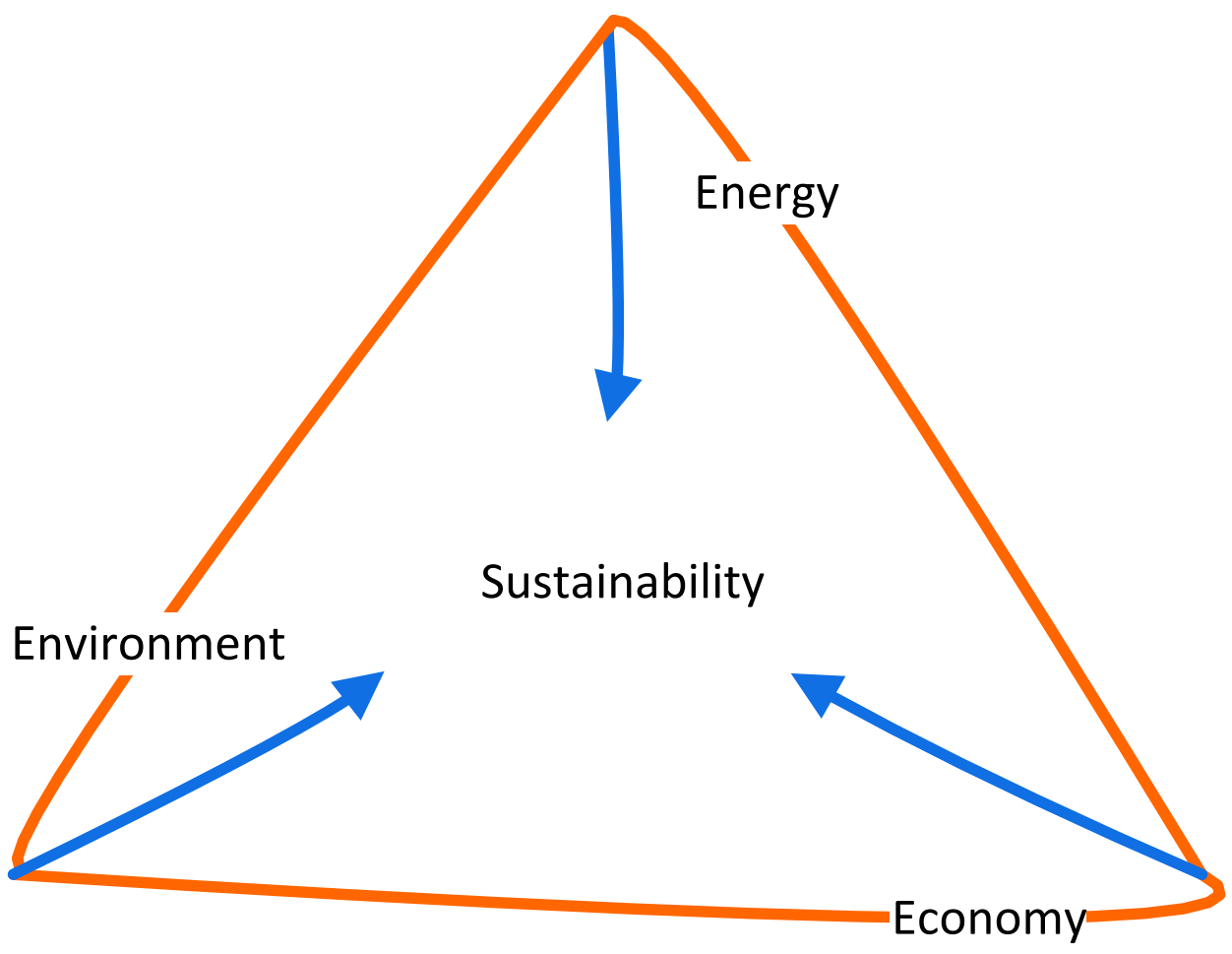

**Fig. 1.** Framework for analysing sustainability of energy systems

# 3. Challenges in China’s energy transition, industry decarbonization and sustainable development

## 3.1 Overall energy development

### 3.1.1 Energy consumption and supply

China is by far the largest primary energy consumer in the world. The primary energy consumption in China experienced a constant increase in the past decade [33]. As shown in the Fig. 2, the energy consumption in 2020 reached 4.56 Btce which is 25% higher than that of in 2010. In addition, fossil fuels still dominate in the energy consumption, and there was no remarkable improvement in the past decade, e.g. 92% of primary energy from fossil fuels compared to 95% in 2010. China’s energy consumption still relies heavily on coal, although the share of coal in total energy consumed has been improved to 62% in 2020 from 73% in 2010. In contrast, the share of crude oil remained at around 20% in the past decade. Natural gas consumption increased dramatically with the amount being twice that of in 2010. However, the overall share is still limited, standing at 9.2% in 2020.

Regarding energy supply, China’s energy sources heavily depend on imports. Fig. 3 shows the imports of main fossil fuels to China from 2010 – 2020. Despite accounting for 13% of world’s coal reserves, China has been a coal exporter prior to 2009 and overtaken Japan as the world’s top coal exporter since 2011 [34]. The net import of coal peaked in 2016, reaching 224.4 Mtoe which is 1.6 times that of in 2010 [34]. Afterwards, it decreased to 32.5 Mtoe in 2010. The external dependence of coal ranged from 1.7% to 11.7% in the past decade. In contrast, China’s economy heavily depend on crude oil and natural gas imports due to the massive demand growth and limited domestic supply. Net oil import almost doubled in the past decade, reaching 506.8 Mtoe in 2020. The external dependence raised from 56.6% in 2010 and ended up at 72.2% in 2020. Natural gas import grew 10 times in the past decade from 10.5 Mtoe in 2010 to 117.6 Mtoe in 2020. As a result, its external dependence raised to 41.3% in 2020.

The fossil energy-dominant energy consumption pattern and high external dependence of fossil energy has threaten the security of long-term and sustainable energy supply. For example, despite the large crude oil import each year, China’s strategic crude oil reserves is less than 60 days which is far below the threshold of a 180-day security reserve [35]. Therefore, it is difficult to maintain a sustainable energy supply, especially under an emergency. In addition, despite the diversity of oil and gas imports, the main importers are from the Middle East, North Africa, Asia Pacific and Russia [34]. Furthermore, the transportation corridors are relatively limited. For example, more than 70% of oil transportation passes through the Strait of Hormuz and the Strait of Malacca, which is highly subject to geopolitical risks [36].

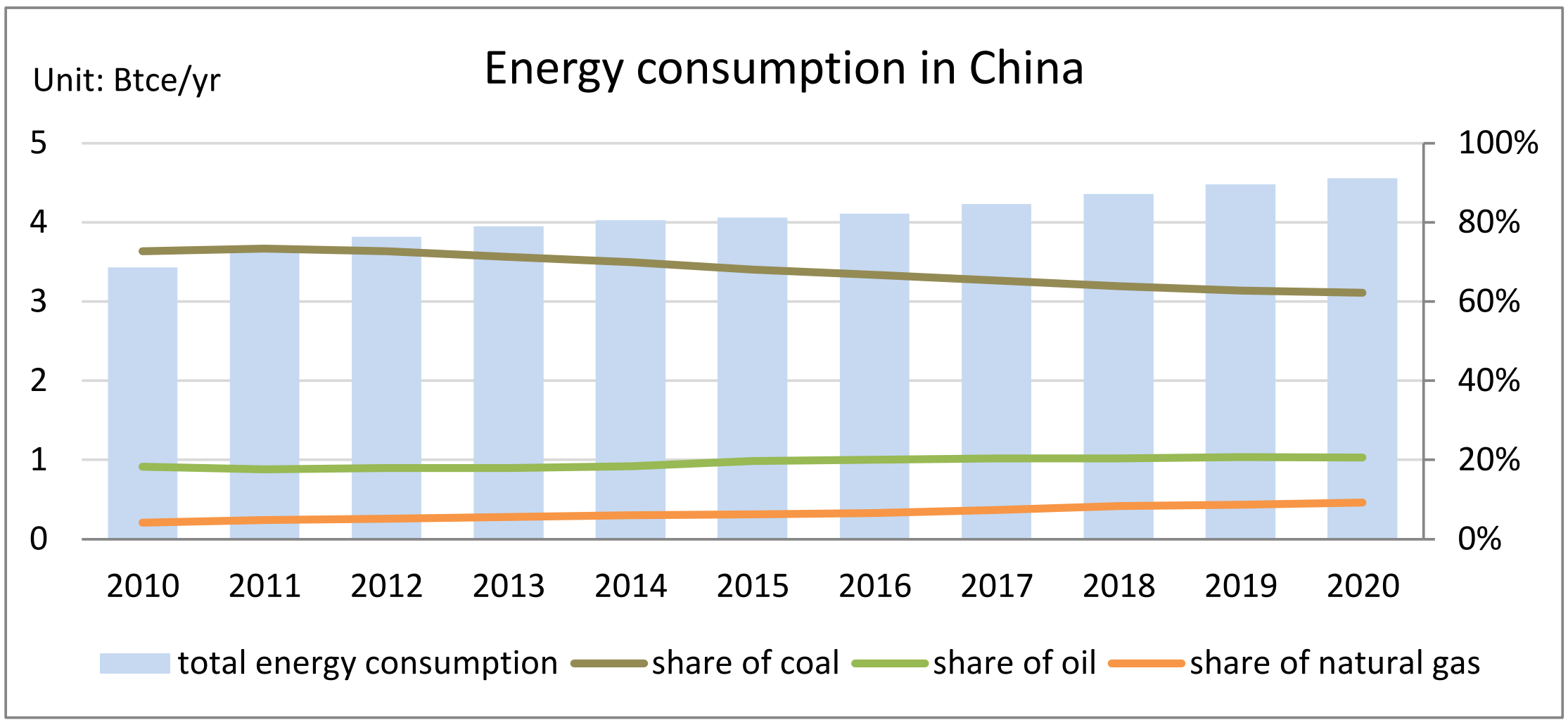

**Fig. 2.** China's energy consumption from 2010-2020

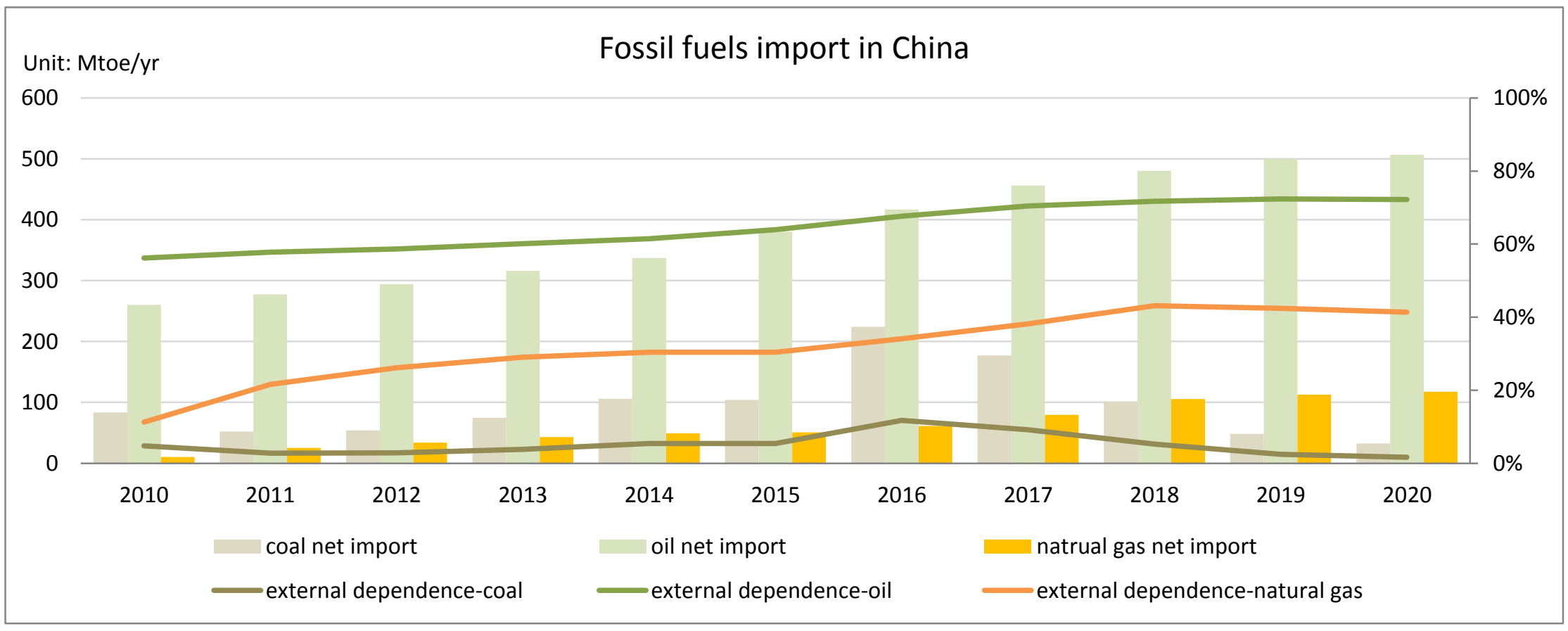

**Fig. 3.** Import of fossil fuels of China from 2010-2020

### 3.1.2 Renewable energy development

As a result of the coal-dominant primary energy supply and steady growth in energy consumption, China has been the world's largest CO2 emitter [34, 37]. The government targeted to phase out fossil fuel use by increasing the share of renewable energy in total energy consumption. The aggregate wind and solar generation capacity reached over 500 GW by 2020 accounting for 23% of total generation capacity in China [38]. Geographically, the installed capacity is mostly distributed in the north, northwest and northeast of China (also known as the Three

North regions) with the share being 60% by 2020, as shown in the Fig. 4, thanks to the abundant renewable resources in these regions [38]. In contrast, economic clusters are generally located in the east and south of China, as indicated in Fig. 5 which shows China’s gross domestic product (GDP) by region in 2020. Therefore, the energy condition in China presents an obvious spatial mismatch between renewable energy supply and potential demand. This has resulted in a sharp decline of utilization and a growing divergence between installed capacity and actual power generation [4]. For example, renewable energy excluding hydro-power only supplied 5.3% of overall energy demand in 2020 [34]. Therefore, large-scale and long-distance renewable energy distribution is fundamental to facilitating the energy transition in China.

Power system extension is by far the centric solution to address energy distribution. By 2020, 20 ultra-high voltage lines were built to support power transmission to load centers, which is far from sufficient considering the fast expansion of renewable power bases in recent years [39]. Meanwhile, the intermittent power aroused challenges for grid integration that the overall operating rate of these lines was lower than 40% in 2019 [40].

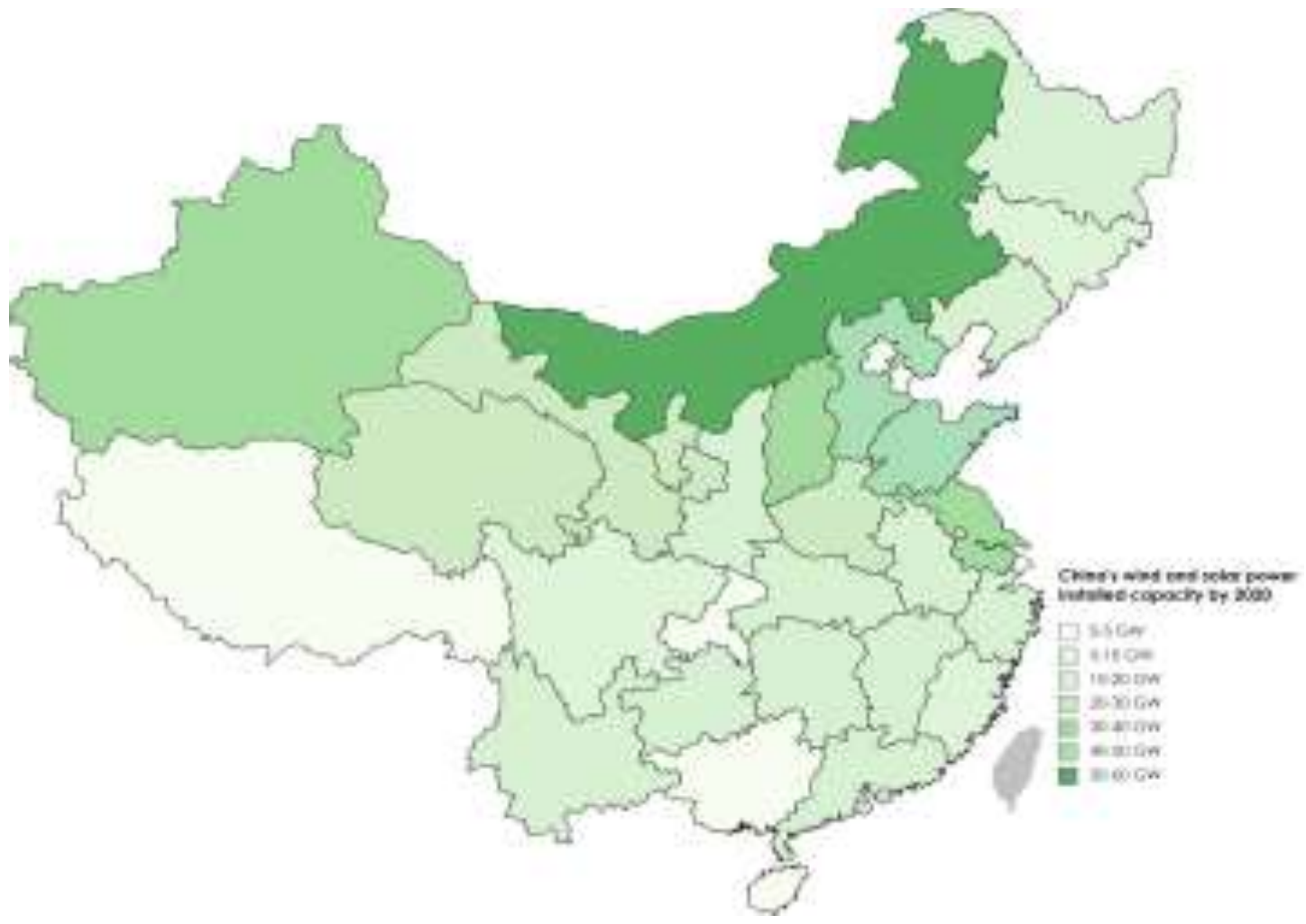

**Fig. 4.** China’s aggregate wind and solar installed capacity by region in 2020

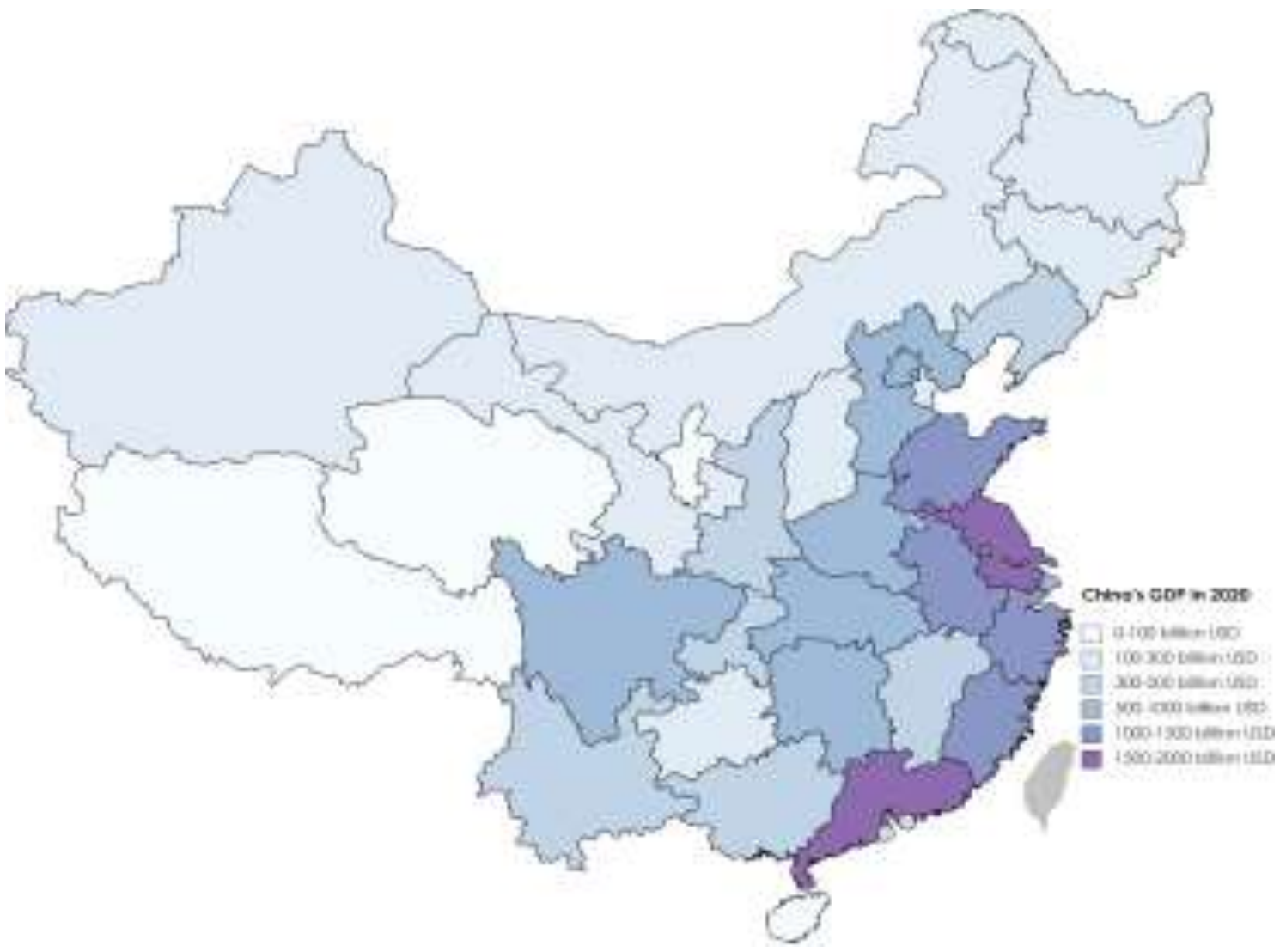

**Fig. 5.** China's gross domestic product (GDP) by region in 2020

## 3.2 Industry decarbonization

### 3.1.1 Overall carbon emission

China is now the world's top CO2 emitter, accounting for around 30% of the global CO2 emission [41]. The CO2 emission in China raised significantly from since 2020 (~3 Gt) and reached 10.38 Gt by 2020, mainly due to the dramatic expansion in manufacturing after joining the WTO [41]. Fig. 6 shows the breakdown of CO2 emission, power and heat consumption in China by sector, where data in 2019 is applied to avoid impacts of covid-19 [42-44].

The overall emission amount in 2019 reached 10.9 Gt, in which, power and heat supply toped other sectors, accounting for 51.7% of the overall China's CO2 emission. The industry contributed 33.9% of overall emission, which is around 3.7 Gt CO2 emission in 2019. Besides, other sectors in total accounts for around 14.4% of CO2 emission. Although China has ranked the first in the world in terms of the number of vehicles (~253.8 M units in 2019), the share of CO2 emission from the mobility sector only accounted for 6.8% in 2019 [45]. The breakdown of power and heat consumption by sector is also analysed to further investigate the causes for the enormous CO2 emission from power and heat supply. Energy consumed in heating accounts for around 13% of the overall energy consumption in both power and heat supply [33]. Therefore, most of CO2 is emitted from power generation process. As also shown in Fig. 7, around 68.2% of power in China in 2019 is consumed by the secondary sector, namely the industry sector [46]. Similarly, around 70% of heat is consumed by the industry sector [47]. Combined with the direct emission from the industry, this indicates that around 70% of CO2 emission in China is caused by the industrial production. As a result, the industry decarbonisation, incl: decarbonisation in production processes and energy conservation, is essential in achieving carbon reduction goals in China.

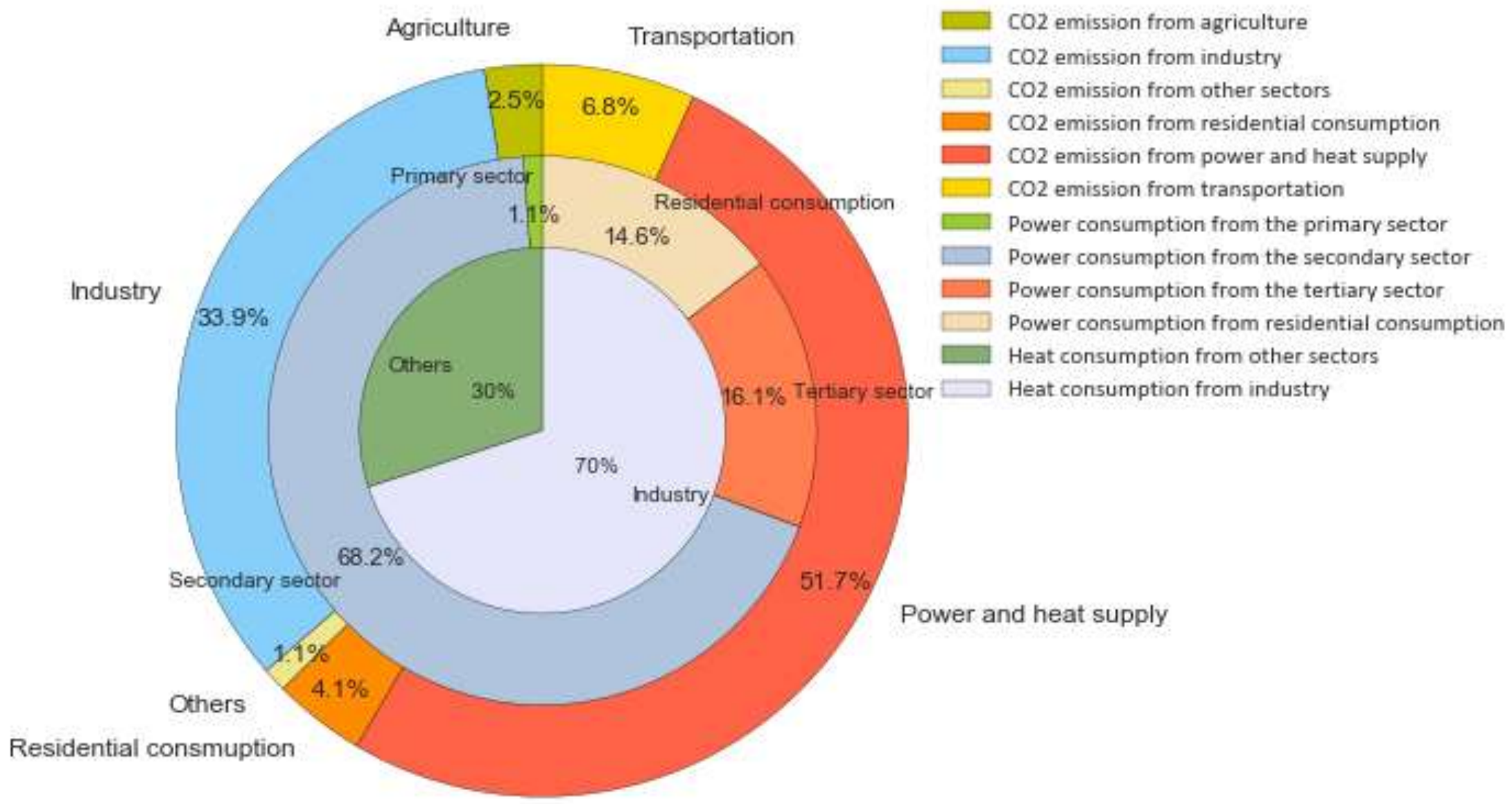

**Fig. 6.** China's CO2 emission, power and heat consumption by sector in 2019

### 3.1.2 The traditional ammonia industry

China has become the world's largest producer of hydrogen with the annual production of around 20 million tons [48]. About 45% of hydrogen produced is used for producing ammonia which is mainly consumed for agricultural and industrial uses [49]. Large amounts of ammonia (~75% in 2017) is used in manufacturing synthetic fertilizers, such as: urea and ammonium bicarbonate [50]. Industrial ammonia (~14% in 2017) is mainly used for producing chemical products, such as: nitrate, acrylonitrile, etc., some of which are used to further produce other chemical products [50].

The ammonia industry in China mainly went through three phases. The industry started at around 1950 to meet the demand from agriculture sector as a direct or indirect fertilizer. The production efficiency is low due to the low facility capacity which is around 40-120 kt/yr [51]. The situation was improved during 1970-1980 when larger size of facilities were imported. Meanwhile, the demand of fertilizers grew dramatically with the development in agriculture sector. To address the increasing gap between supply and demand, large amounts of ammonia was imported which accounts for 25% of global ammonia supply. The ammonia industry experienced a fast growth in the second phase that starts at about 1980 after the economic reform and open-door policy launched. With more governmental policy support, key technologies were improved and out-dated facilities and small producers were replaced by producers with more efficient production. As a result, production capacity was able to meet the domestic demand by 2000. The industry was further boosted after China became a member of the World Trade Organization (WTO). The overall production reached around 55 Mt in 2012, representing one-third of global total production [50]. About 49% production capacity is owned by large enterprises which are 74 out of 496 ammonia producers in 2010 [50]. The third phase featured a further adjustment after a rapid development in the industry. Ammonia industry experienced overcapacity, due to the saturated domestic demand and low cost ammonia supply overseas. As a result, many ammonia producers were phased out, and production capacity encountered a sharp decline from 2015-2017 [33], as shown in the Fig. 7.

China is now the world's largest ammonia producer, with the production capacity widely distributed in each region, as shown in the Fig. 8, especially in the north, southwest, mid-south and east of China, such as: in Shandong, Henan, Shanxi, Hebei, Hubei and Jiangsu province [33]. The industry consumes the most energy in China's chemical industry, as its production process is very energy intensive. As shown in Table 1, coal-based ammonia production through the coal gasification (CG) process is dominant in China due to the rich coal reserves [51]. In contrast, about 72% of ammonia is produced by using natural gas (NG) via the steam reforming (SMR) process on a global scale. NG-based ammonia production in China is mainly distributed in Sichuan and Xinjiang province due to more NG reserves in these regions [50]. As a result, ammonia industry top other chemical industries in terms of CO2 emission, and also emits other pollutants such as sulphur dioxide (SO2), nitrogen oxides (NOx), etc. These have significantly contributed to environmental problems in China [50].

**Table 1**

Proportion of energy feedstocks for ammonia production in China

| Year | Coal | Natural gas | Coke oven gas | Oil | Others |
|---|---|---|---|---|---|
| 1987 | 67.4% | 18.5% | 0.1% | 14% | 0.4% |
| 1997 | 69.6% | 19.9% | 0.1% | 9.8% | 0.5% |
| 2007 | 79.2% | 18.4% | 1% | 1.1% | 0.3% |
| 2017 | 75.7% | 20.6% | 3.1% | 0.5% | 0.1% |

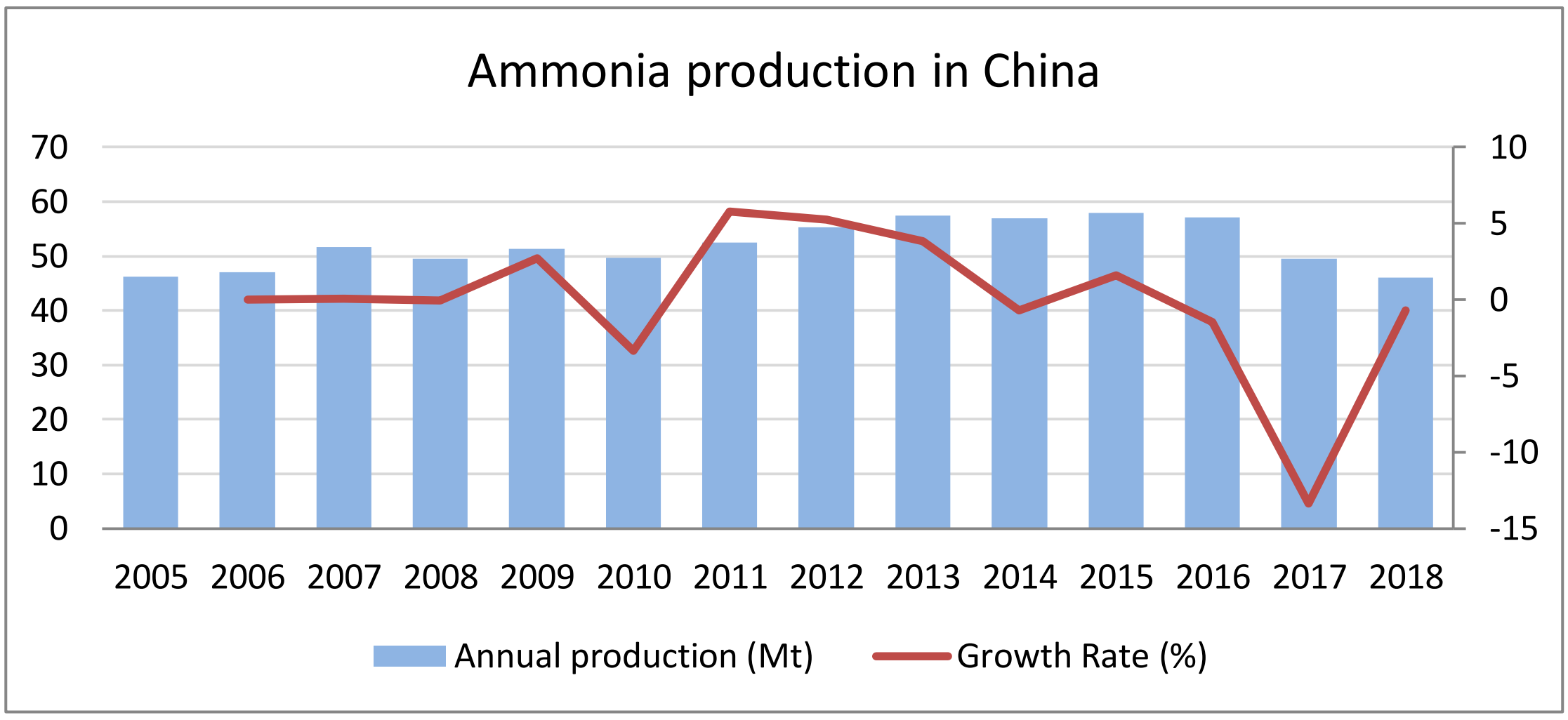

**Fig. 7.** China's ammonia production from 2005-2018

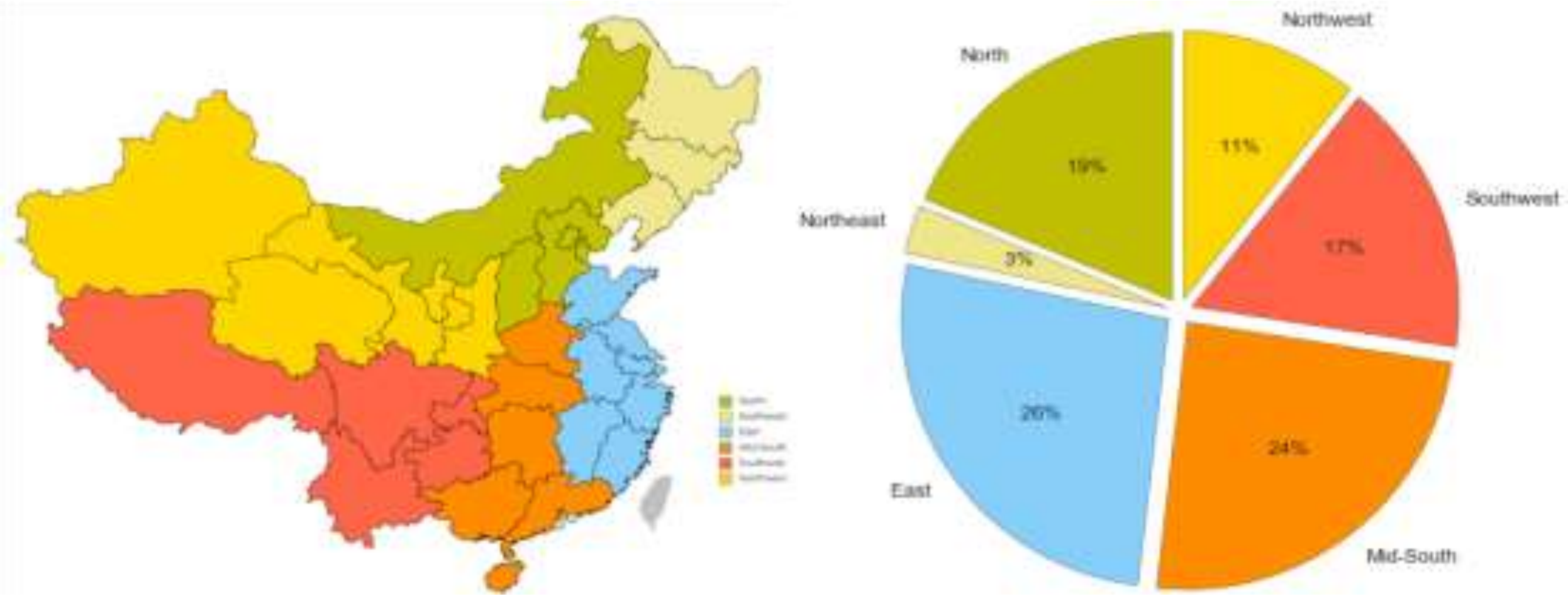

**Fig. 8.** China's ammonia production by region in 2018

## 3.3 Regional sustainable development

Sustainable development concerns not only socio-economic growth, but also energy use and environmental impacts [52]. Compared to the traditional approaches, the green productivity, or known as green total-factor productivity (GTFP), is increasingly paid attention as a measurement of synergistic performance between economic growth and environmental costs [53]. The approach of data envelopment analysis (DEA) has been widely applied in assessing total-factor productivity, as it can compare the technical efficiency between decision-making units (DMUs) with multiple and similar types of inputs and outputs [54]. We apply a constant returns to scale (CRS) DEA model to assess the regional GTFPs in China. The objective function and associated constraints of the *i*th DMU are formulated in Eq.(1)-(4), where $\vartheta_i$ denotes the efficiency of the DMU *i* ranging from 0 – 1; $X_i$ and $Y_i$ are the inputs and outputs of the DMU *i*, respectively; $X_j$ and $Y_j$ are the inputs and outputs of the DMU *j*, respectively; $\lambda_j$ is the weight of the DMU *j*.

Fig. 9 shows the input-output framework of the DEA model. Labor force and capital stock are considered as socio-economic input factors. In addition, energy consumption and carbon emission are also considered as energy and environmental costs in measuring the green productivity. Gross domestic product (GDP) is considered as the output of the DEA model. Regional GTFPs in 2019 are measured, considering the impact of covid-19 from 2020-2022 in China. Table 2 shows parameters applied in the assessment, in which, data is collected from 31 administrative areas in China and integrated into six regions, including: north, northeast, east, mid-south, southwest and northwest of China. CO2 emission of Tibet in 2019 is estimated based on the level in 2014 and average growth rate. Regional capital stock is estimated on the basis of the level in 2000 estimated by Zhang et al. and the perpetual inventory method [55]. As shown in Eq.(5), the perpetual inventory method is applied to calculated the capital stock of a following year, where $K_{i,n}$ and $K_{i,n+1}$ denote the capital stock of year *n* and *n+1* in region *i*, respectively; $I_{i,n+1}$ denotes fixed capital stock of year *n+1* in region *i*; $\delta$ is the depreciation rate which is set as 9.6% as Zhang proposed [56]. CO2 emission of Tibet in 2019 is missing in China's Carbon Emission Accounts and Datasets, is thus estimated based on the emission level in 2014 and annual average growth rate of China from 2014-2019 [57, 58]. Besides, energy intensity and carbon intensity are also calculated based on Eq.(6) and Eq.(7), respectively, where $E_{i,n}$ and $EI_{i,n}$ denote energy consumption and intensity of region $i$ in year $n$, respectively; $C_{i,n}$ and $CI_{i,n}$ denote CO2 emission and carbon intensity of region $i$ in year $n$, respectively; $G_{i,n}$ is the GDP of region $i$ in year $n$.

$$\min \theta_i, \ \ s.t. \tag{1}$$

$$\sum_j X_j \lambda_j \leq \theta X_i, \ \ \forall j \in (1,2, \dots, \mathrm{M}\} \tag{2}$$

$$\sum_j Y_j \lambda_j \geq Y_i \ , \forall j \in (1,2, \dots, \mathrm{M}\} \tag{3}$$

$$\lambda_j \geq 0, \ \ \forall j \in (1,2, \dots, \mathrm{M}\} \tag{4}$$

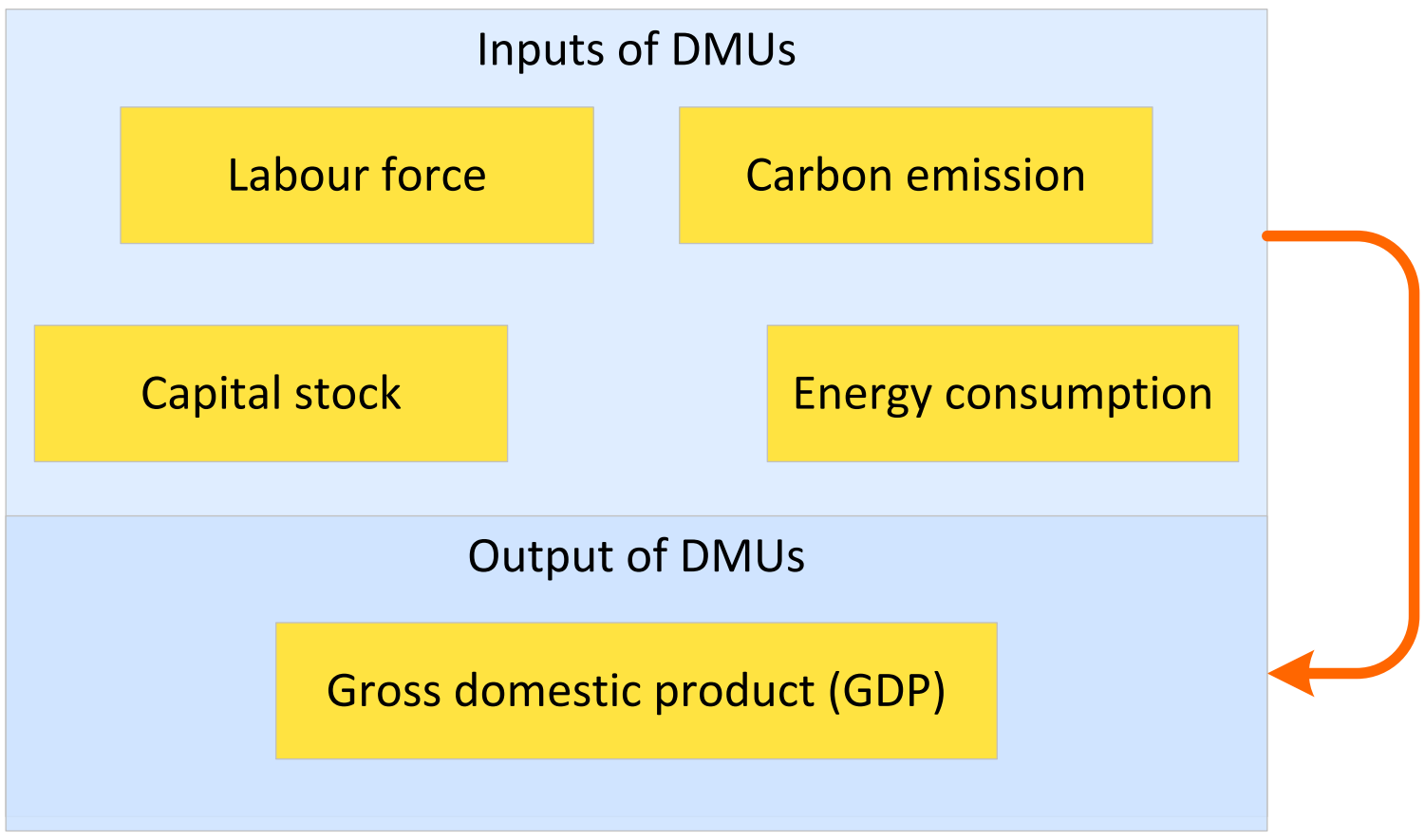

**Fig. 9.** Framework of the DEA model

**Table 2**

Parameters for evaluating regional GTFP of China in 2019

| Parameter (2019) | Region | Value | Unit | Source |
|---|---|---|---|---|
| Energy consumption | North | 943.50 | M tce/yr | [33, 59] |
| | Northeast | 408.49 | | |
| | East | 1452.58 | | |
| | Mid-South | 1032.93 | | |
| | Southwest | 515.58 | | |
| | Northwest | 521.88 | | |
| Labour force | North | 87.28 | M p/yr | [33] |
| | Northeast | 50.75 | | |
| | East | 244.29 | | |
| | Mid-South | 220.64 | | |
| | Southwest | 114.33 | | |
| | Northwest | 54.62 | | |
| Capital stock | North | 2290.21 | B USD | [33] |
| | Northeast | 795.42 | | |
| | East | 4428.91 | | |
| | Mid-South | 3433.75 | | |
| | Southwest | 1707.82 | | |
| | Northwest | 1009.98 | | |
| CO2 emission | North | 2521.59 | Mt CO2/yr | [42, 57, 58] |
| | Northeast | 1015.26 | | |
| | East | 3244.51 | | |
| | Mid-South | 2001.62 | | |
| | Southwest | 924.63 | | |
| | Northwest | 1180.19 | | |
| GDP | North | 1697.42 | B USD | [33] |
| | Northeast | 717.84 | | |
| | East | 5363.89 | | |
| | Mid-South | 3919.61 | | |
| | Southwest | 1257.92 | | |
| | Northwest | 783.19 | | |

$$K_{i,n+1} = I_{i,n+1} + (1-\delta)K_{i,n} \quad (5)$$

$$EI_{i,n} = E_{i,n}/G_{i,n} \quad (6)$$

$$CI_{i,n} = C_{i,n}/G_{i,n} \quad (7)$$

Fig. 10 shows the results of regional GTFPs, energy intensities and carbon intensities of China in 2019. Both east and mid-south of China reach the optimal productivity with GTFP being 1. In addition, energy and carbon intensity in mid-south of China are the lowest in all the six regions, being 7.32 Btu/USD and 0.51 kg CO2/USD, respectively. East of China is slightly higher than the Mid-South in term of energy and carbon intensity, but much lower than that of in other regions. This indicates that east of China is more efficient in converting labour and capital factors into economic outputs compared to the Mid-South, which can attribute to the technological, organizational factors, etc. In general, both two regions have less energy-intensive and pollutant industry and higher overall productivity. In contrast, GTFPs in the three North regions are low, and energy and carbon intensity are higher than the rest of regions. Northwest of China obtains the lowest productivity in China, with its GTFP being 0.65. Further, its energy and carbon intensity are also the highest, being 18.51 Btu/USD and 1.51 kg CO2/USD, respectively. It is also the case for the north and northeast of China, where energy and carbon intensity are also high, but GTFPs are relatively higher than that of the Northwest, being 0.89 and 0.75, respectively. It indicates that there is a large-scale energy-intensive and pollutant industry in the Three North regions, and the Northwest especially needs to improve technology, management and industrial structure, due to the lowest GTFP. The GTFP of southwest of China is slightly higher than that of in the Northwest, however, its energy and carbon intensity are much lower, being 11.38 Btu/USD and 0.74 kg

CO2/USD, respectively. This indicates there is less heavy industry in this region, but lacks of technological factors, etc.

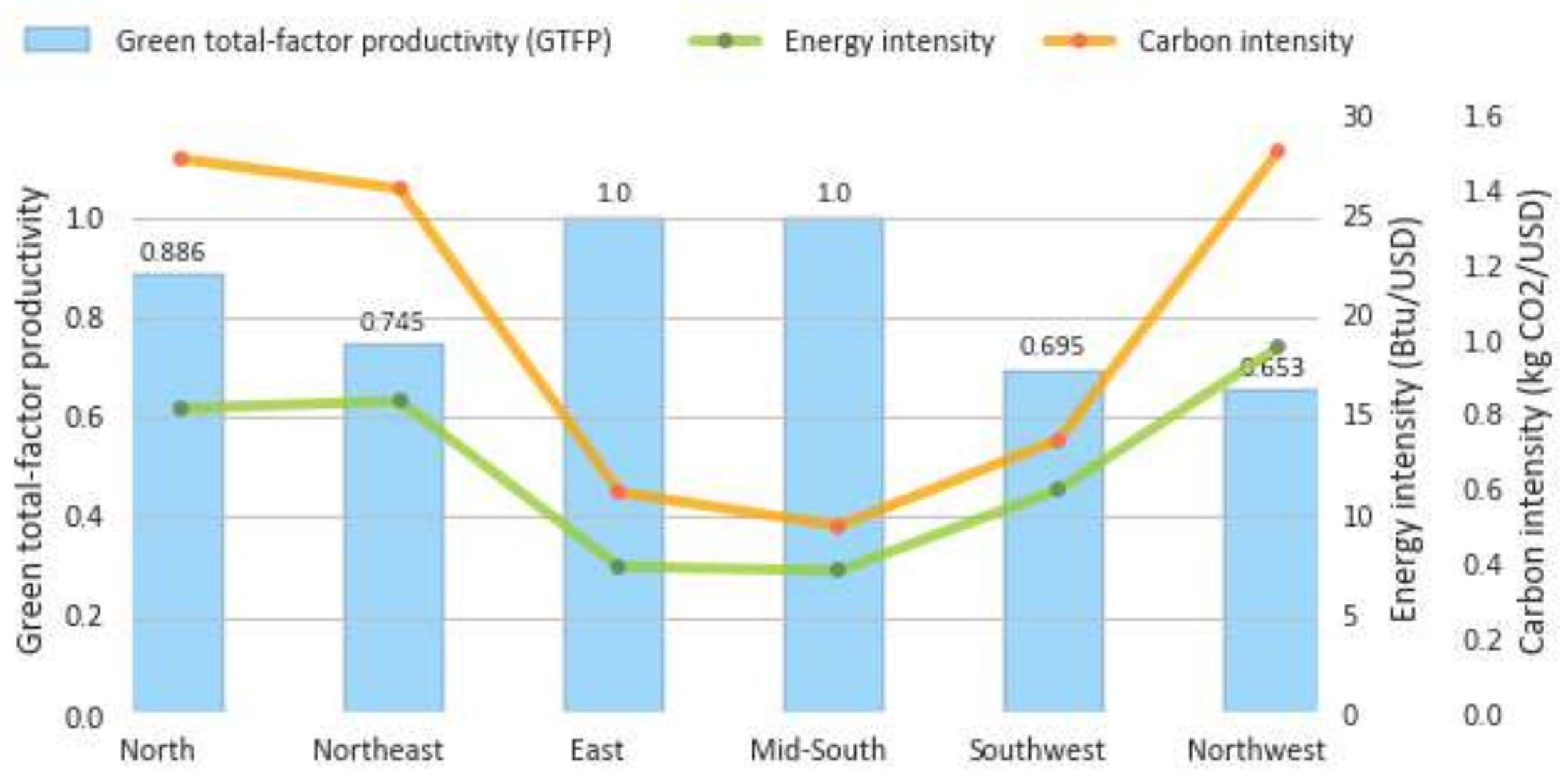

**Fig. 10.** GTFP and energy and carbon intensity in the six administrative regions of China in 2019

# 4. Green ammonia in meeting China's sustainability challenges

Confronting challenges towards a sustainable development in China, the further development of renewable energy is a necessity. The settlement of administrative barriers and introduction of market-based solutions are crucial to increase the share of renewable energy [60]. Meanwhile, a concerted effort from hydrogen is required for large-scale energy storage and distribution [6]. Since green ammonia is key in the hydrogen-based energy transition as a potential hydrogen carrier and clean fuel, the sustainability of developing the green ammonia industry are comprehensively discussed in this section.

## 4.1 Green ammonia as a hydrogen carrier

The economic feasibility of transporting energy as ammonia is examined by comparing the delivery and storage costs with the major options of liquid hydrogen and pipeline. The related levelized costs are calculated based on Eq.(8)-(9), where $Exp_n$ denotes the total expense in year $n$, which consists of investment cost ($Capex_n$) and operating cost ($Opex_n$), including: maintenance cost and fuel cost. $E_n$ is the energy delivered or stored per year $n$, and $dr$ is the discount rate. The applied parameters are listed in Table 3, in which, a median conversion rate of ammonia reform is applied based on the literature [61, 62].

Fig. 11 shows breakdown of hydrogen delivery costs for these options by volume with a 500 km distance. The options of ammonia and liquid hydrogen are less influenced by volume. Delivery cost of ammonia remains at around 1.4-1.5 USD/kg H2 in all cases, and is dominated by ammonia production cost and decomposition cost. The sheer transportation cost only accounts for around 5% of the overall cost. In contrast, although pipeline appears to be the ultimate solution, hydrogen delivery is influenced dramatically by volume due to the high capital and operating expenses required. The effect of delivery distance is also examined for transporting hydrogen from 500 to 3000 km, as shown in Fig. 12. Similarly, hydrogen delivery by pipeline is influenced dramatically by volume with the cost ranging from 0.6-3.3 and 0.8-5.1 USD/kg H2 for delivering 100 and 50 kt hydrogen per year, respectively. Liquid hydrogen and ammonia are more flexible options which have similar costs and limited influences by distance. The transportation will be more efficient if ammonia can be directly used. In this case, ammonia reform is excluded, and the overall cost can reach around 0.8-1.2 USD/kg H2, as the light blue line shows.

Fig. 13 shows costs of hydrogen storage by time in the form of ammonia and liquid hydrogen. Underground storage which is normally used for pipeline transportation is not presented due to a lack of data. The case of transporting 100 kt hydrogen per year is examined assuming 20% of the total volume is stored for short-term and seasonal supply (i.e. 30 and 150 days, respectively) considering the proportion for natural gas is at least 15% [63]. Storage costs for longer periods are also assessed to examine the potential for future long-term energy reserve, as strategic petroleum and gas reserves of today. In general, hydrogen storage as ammonia is little impacted by storage length. Storage costs range from 0.6-0.7 USD/kg H2 which is 3-4 times less than that of liquid hydrogen, due to mild storage conditions. Liquid hydrogen storage experience a stable increase in cost over time due to the incremental energy for extreme cryogenic condition required. In addition, ammonia defeats liquid hydrogen in terms of safe transportation and long-term storage due to much lower flammability [7]. Therefore, these advantages enable transporting hydrogen as ammonia a flexible, economical and safe option. Since infrastructure for ammonia is already in place due to a century of use in agriculture, the development of green ammonia supply chains and direct use of ammonia are proposed to avoid large uncertainties and significant initial investment in the early stage.

$$LCOE = \sum_n \left(\frac{Exp_n}{(1+dr)^n}\right) / \sum_n \frac{E_n}{(1+dr)^n} \tag{8}$$

$$Exp_n = Capex_n + Opex_n \tag{9}$$

**Table 3**

Parameters for evaluating hydrogen transport and storage costs

| Parameter | Value | Unit | Source |
|---|---|---|---|
| Weighted average cost of capital | 8 | % | [23] |
| Capital cost of ammonia plant | 714 | USD/t NH3 | [64] |
| Capital cost of ammonia reformer (10 kt/yr) | 354 | USD/t NH3 | [61] |
| Capital cost of ammonia reformer (30 kt/yr) | 267 | USD/t NH3 | [61] |
| Capital cost of ammonia reformer (50 kt/yr) | 234 | USD/t NH3 | [61] |
| Capital cost of ammonia reformer (100 kt/yr) | 196 | USD/t NH3 | [61] |
| Capital cost of ammonia tanker | 30,000 | USD | [65] |
| Capital cost of ammonia vessel | 808 | USD/t | [62] |
| Capital cost of liquefier (10-30 kt/yr) | 8,219 | M USD/t H2 | [66] |
| Capital cost of hydrogen liquefier (50 kt/yr) | 7,397 | M USD/t H2 | [66] |
| Capital cost of hydrogen liquefier (100 kt/yr) | 6,301 | M USD/t H2 | [66] |
| Capital cost of hydrogen vaporizer (10 kt/yr) | 20 | K USD | [65] |
| Capital cost of cyro-tank | 1000 | USD/cu m | [65] |
| Capital cost of hydrogen pipeline (10 kt/yr) | 531 | K USD/km | [67] |
| Capital cost of hydrogen pipeline (30 kt/yr) | 599 | K USD/km | [67] |
| Capital cost of hydrogen pipeline (50 kt/yr) | 665 | K USD/km | [67] |
| Capital cost of hydrogen pipeline (100 kt/yr) | 833 | K USD/km | [67] |
| Conversion rate of ammonia synthesis | 95 | % | [68] |
| Conversion rate of ammonia reform | 95 | % | [61, 62] |
| Energy for heating and pressurization in ammonia production | 0.6 | MWh/t | [69] |
| Energy for decomposing ammonia | 1.4 | MWh/t | [62] |
| Energy for ammonia cooling | 0.19 | MWh/t | [62] |
| Energy for ammonia storage | 5.6 | kWh/t/d | [62] |
| Energy for hydrogen liquefaction | 10 | MWh/t | [66] |
| Energy for hydrogen regasification | 0.06 | kWh/t | [70] |
| Energy for hydrogen pipeline transport | 0.26 | MWh/t/100 km | [62] |
| Evaporation in ammonia transport | 0.024 | %/d | [71] |
| Evaporation in liquid hydrogen transport | 0.2 | %/d | [72] |
| Leakage in hydrogen pipeline transport | 0.05 | %/1000 km | [73] |

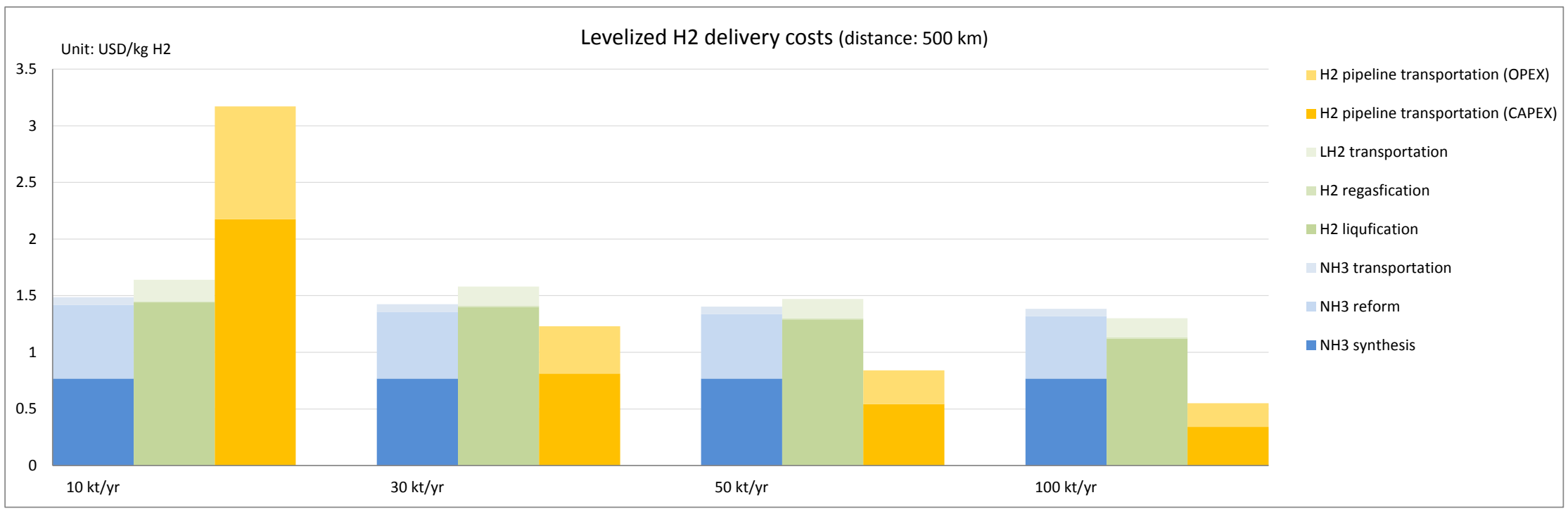

**Fig. 11.** Impacts of transport volume on hydrogen delivery cost

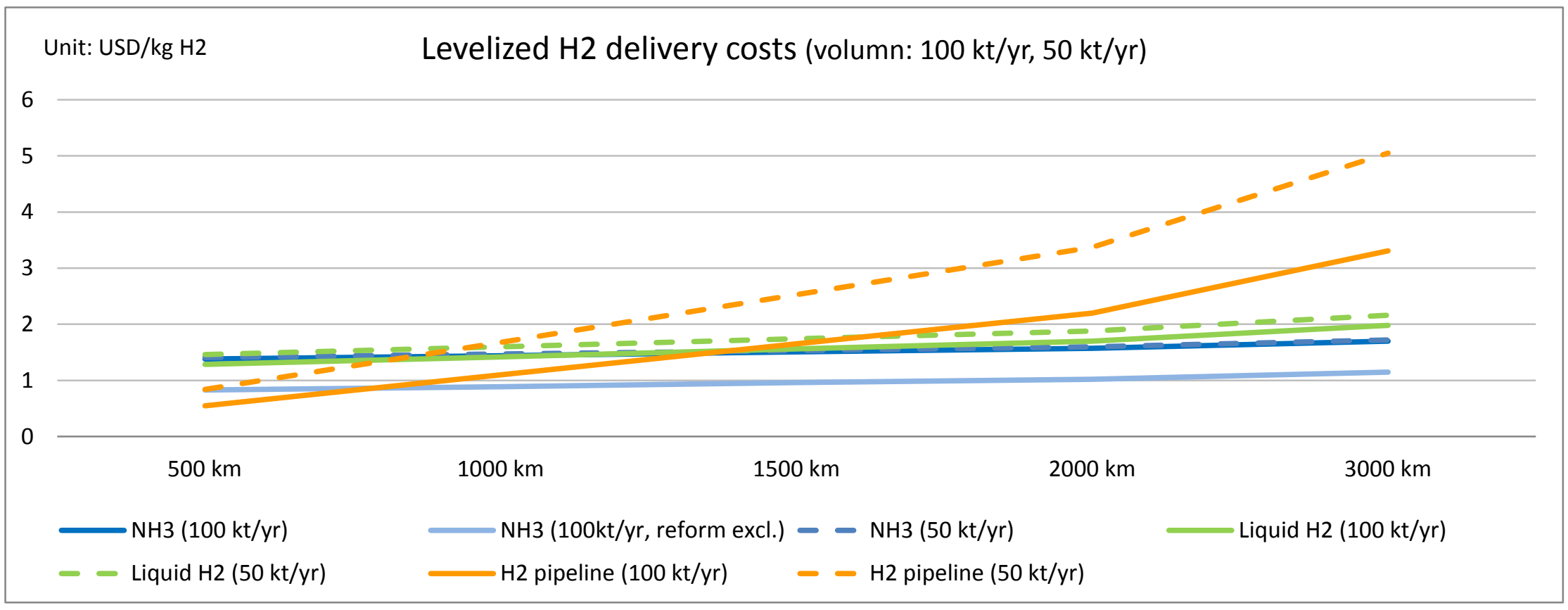

**Fig. 12.** Impacts of transport distance on hydrogen delivery cost

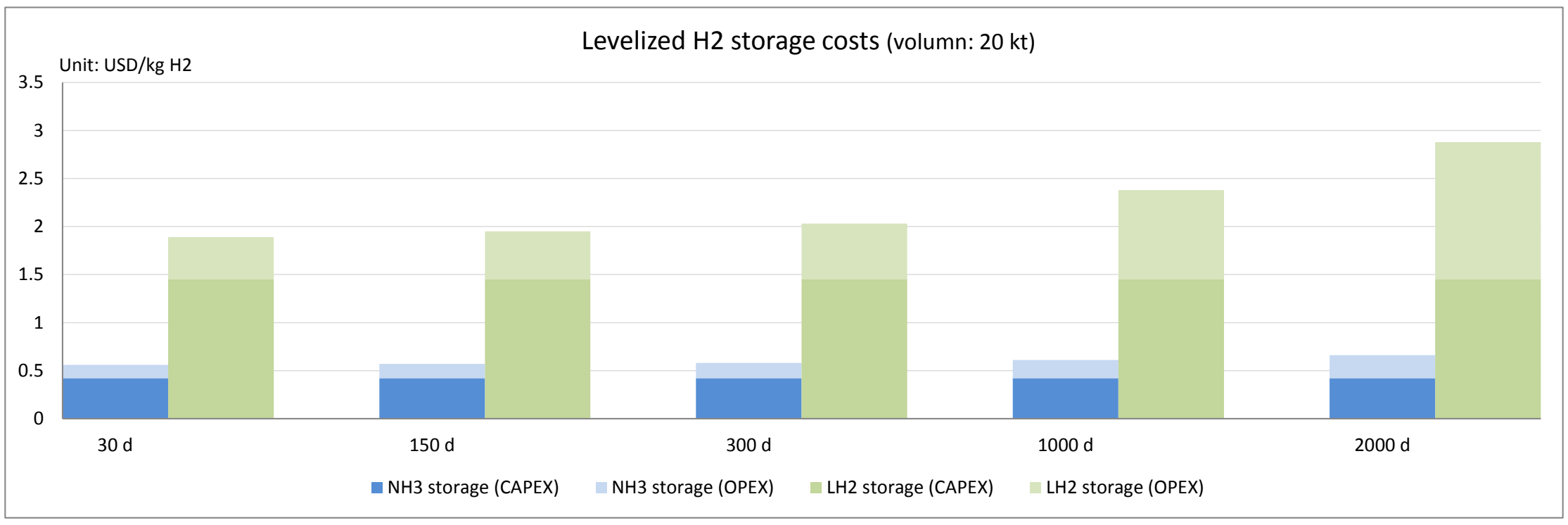

**Fig. 13.** Impacts of storage time on hydrogen storage costs

## 4.2 Green ammonia as a clean fuel and fertilizer

Green ammonia is emerging as a clean fuel with co-firing ammonia with coal in power generation as an important application to decarbonize the power sector and provide sufficient demand. The mixed fuel cost and related power generation cost are evaluated based on average thermal coal trading price from 2012-2021 and green ammonia price, in which, green ammonia price is estimated with production cost and gross margin in the works [23, 24]. Scenarios with different co-firing rate ranging from 0-20% are defined in Table 4. Co-fired power generation cost and fuel cost are evaluated based on Eq.(10)-(11), where $LCOE^m$ denotes levelized cost of electricity in the co-firing power generation; $FC^m$ denotes consumption of the mixed fuel for power generation per unit; $FE^m$ denotes mixed fuel cost; $EP^c$ denotes share of coal cost in power generation; $LCOE^c$ denotes levelized cost of coal power; $FE^{am}$ denotes cost of green ammonia; $FE^c$ denotes cost of thermal coal; $FR^m$ denotes co-firing rate. Parameters applied for evaluating costs of fuel ammonia are shown in Table 5.

Fig. 14 shows fuel and electricity costs and carbon emission intensity for ammonia co-firing. Both electricity and fuel costs undergo a stable increase with ammonia co-firing scaled up. Since currently green ammonia remains expensive, at 3% co-firing, the mixed fuel cost grows by 23.5%, which leads to a 17% increase in levelized cost of electricity. However, the fuel cost is still lower than the maximum coal price in this period (2012-2021) which is 195.4 USD/tce, and much lower than the average natural gas price which is 262.7 USD/tce. Around 25.1 kg CO2 emission is reduced per MWh of power generated. A 5% ammonia co-firing resulted in a 39.2% increase in fuel cost and a 29.4% increase in power generation cost. The fuel cost reaches 213.5 USD/tce which is slightly higher than the maximum coal price, but still much lower than the average natural gas price. Carbon emission intensity is reduced by 41.9 kg CO2/MWh. Mixed fuel cost rises by 2.6 times and power generation cost reaches 150.3 USD/MWh when co-firing rate is 20%, compared to the base scenario without co-firing considered. However, the mixed fuel cost in this case is still much lower than the maximum gas price.

Besides, the comparison of mixed fuel cost and that of natural gas does not consider efficiency of thermal power units. In practice, despite of the higher power generation efficiency of natural gas-fired units (~52%) than that of coal-fired units (~38%), the integrated efficiency (incl. coal and heat generation) is lower than that of coal-fired units (~57.6% vs. ~75.7%) in China [74]. Considering the high proportion of cogeneration (~62%) in China's coal power plants [74], the mixed fuel cost is further economic. Further, the cost calculation is based on the average thermal coal price from 2012-2021. In practice, the fuel cost can be guaranteed and further reduced with long-term contracts with fuel suppliers. Therefore, ammonia co-firing at 3-5% rates, especially 3% rate in the near future can be options for China for low-carbon transition, compared to increasing the share of natural gas in power generation.

So far, grey ammonia produced from fossil fuels is primarily used as a fertilizer in agriculture. The price in China fluctuates between 285-500 USD/t with an average price at around 430 USD/t in the past decade [23, 75]. In contrast, green ammonia price ranges from 700-1400 USD/t worldwide, with the level of 700-800 USD/t primarily reported and estimated [76-78]. Levelized cost of green ammonia is estimated at 720-820 USD/t in eastern Inner Mongolia in the north of China, where enormous availability of renewable resources are located [24]. In the light of this, green ammonia price is currently at least around twice that of grey ammonia in China [24]. Therefore, a moderate increase of the share of green ammonia, similar to ammonia co-firing in power generation, provides a potential pathway to gradually phase out the use of grey ammonia in agriculture.

$$LCOE^m = FC^m * FE^m + (1 - EP^c) * LCOE^c \tag{10}$$

$$FE^m = FE^{am} * FR^m + FE^c * (1 - FR^m) \tag{11}$$

**Table 4**

Definition of scenarios of co-firing ammonia in coal power generation

| Co-firing scenario | Co-firing rate |
|---|---|
| Base case | 0% |
| Case 1 | 3% |
| Case 2 | 5% |
| Case 3 | 10% |
| Case 4 | 15% |
| Case 5 | 20% |

**Table 5**

Parameters for evaluating costs of ammonia co-firing

| Parameter | Value | Unit | Source |
|---|---|---|---|
| Thermal coal trading price (2012-2021) | 142.4-195.4 | USD/tce | [79] |
| LNG trading price (2012-2021) | 419.8-516.4 | USD/tce | [79] |
| Lower heating value of thermal coal | 5500 | kcal/kg | [79] |
| Lower heating value of ammonia | 18.6 | MJ/kg | [80] |
| Green NH3 production cost | 820 | USD/t | [24] |
| Gross Margin of ammonia sales | 5 | % | [23] |
| Coal consumption in power generation | 0.31 | t/MWh | [81] |
| CO2 emission of coal power generation | 838 | kg/MWh | [82] |
| Efficiency loss (3% co-firing) | 1 | % | [82] |
| Efficiency loss (5% co-firing) | 2 | % | [82] |
| Efficiency loss (10% co-firing) | 3 | % | [82] |
| Efficiency loss (15% co-firing) | 4 | % | [82] |
| Efficiency loss (20% co-firing) | 6 | % | [82] |
| Share of fuel cost in overall cost of coal power generation | 70 | % | [83] |

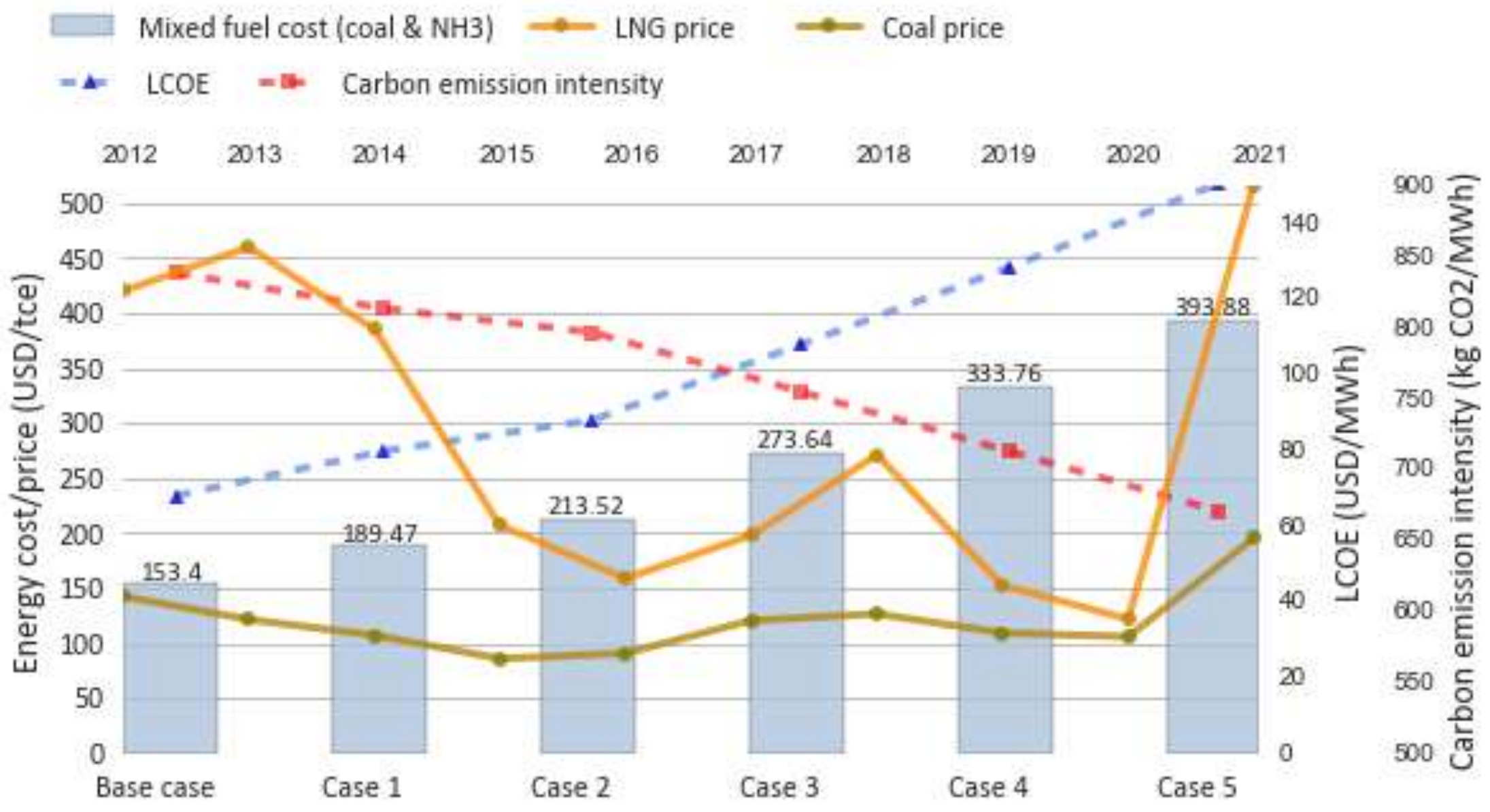

**Fig. 14.** Fuel and electricity costs and carbon emission intensity for co-firing ammonia

## 4.3 Supply and demand of green ammonia

As estimated above, the direct use of green ammonia is more economic especially in the early stage of a hydrogen economy. Therefore, the conventional ammonia industry where ammonia is being used as a fertilizer or an intermediate chemical product is a potential consumer which exists in a large scale in the eastern and southern China [84]. In addition, ammonia to be used as a clean fuel for thermal power generation and shipping are also potential options, which are either taking place or being considered in some countries [68]. Furthermore, the demand from mobility sector is also taken into account, since the development of fuel cell vehicles (FCVs) is considered a high priority in China, and hydrogen is expected to achieve price parity with gasoline by 2025 [85, 86].

The potential green ammonia supply capacity and demand by 2030 are evaluated with parameters shown in Table 6, in which, demands of conventional ammonia and heating oil for shipping, annual use hours of coal power, and its share in thermal power generation in 2030 are estimated based on sources [33, 87]. The related scenarios for supply capacity and demand are defined in Table 7 and Table 8, respectively. Three supply levels are categorized by the share of renewable power (incl. wind and solar power) capacity available for ammonia production with Level 1 concerning a possible case and Level 3 entailing an ideal condition. Potential demand is divided into five levels with penetration rates (PRs) of green ammonia in conventional ammonia sector, power sector and mobility sector rising from Level 1-5. Penetration rate in mobility sector ($PR^{m}$) is defined in Eq.(12), where $UR^{hrs}$ denotes utilization rate of hydrogen refilling stations (HRSs), and $TR^{am}$ denotes the share of hydrogen transported as ammonia in 2030.

$$PR^{m} = UR^{hrs} * TR^{am} \quad (12)$$

Fig. 15 shows the potential supply capacity and demand levels by 2030. The lowest level of demand can be met if 15% of renewable energy is used for green ammonia production. However, the ideal supply level can only meet around 64% of the highest level of demand. This is because of a massive volume of energy required per year for power generation in China, and a slight increase in ammonia co-firing rate results in a dramatic rise in ammonia demand. In contrast to Japan's plan to achieve 20% ammonia co-firing rate in coal power plants by 2030 [88], a 5% co-firing rate requires almost a half of overall wind and solar power capacity in China. A maximum of 3% co-firing rate is more reasonable, which requires about 28% of renewable power generation in China by 2030. The ammonia industry can provide sufficient demand, although it is relatively small compared to the dominant demand from power generation. A 50% transition to green ammonia production consumes about 10% renewable power generation by 2030. In contrast, despite of China's ambitious goals to have 1 million FCVs on the road by 2030 [89], the share of demand from FCVs is relatively insignificant that only 1.6 Mt of ammonia is required annually in the ideal case. The demand of ammonia as shipping fuel is 6.7 Mt per year if 15% of heating oil is replaced, which is higher than that of mobility sector.

**Table 6**

Parameters for evaluating green ammonia supply and demand in 2030

| Parameter | Value | Unit | Source |
|---|---|---|---|
| Wind power capacity in 2030 | 780 | GW | [90] |
| Solar power capacity in 2030 | 840 | GW | [90] |
| Thermal power capacity in 2030 | 1450 | GW | [91] |
| Demand of conventional ammonia in 2030 | 52 | Mt | [33] |
| Demand of heating fuel for shipping in 2030 | 20 | Mt | [33] |
| Annual use hours of wind power | 2246 | h | [33] |
| Annual use hours of solar power | 1163 | h | [33] |
| Annual use hours of coal power | 4000 | h | [87] |
| Proportion of coal power in thermal power | 87 | % | [87] |
| Electrolyser efficiency | 70 | % | [7] |
| Number of HRSs in 2030 | 1000 | ea. | [89] |
| Average capacity of HRSs | 1000 | kg H2/d | [89] |

**Table 7**

Definition of scenarios for green ammonia supply capacity

| Supply capacity scenario | Share of wind and solar power capacity required |
|---|---|
| Level 1 | 15% |
| Level 2 | 35% |
| Level 3 | 65% |

**Table 8**

Definition of scenarios for green ammonia demand

| Demand scenario | PR in ammonia sector | PR in power sector | PR in shipping sector | PR in mobility sector |
|---|---|---|---|---|
| Level 1 | 10% | 1% | 3% | 10% |
| Level 2 | 20% | 3% | 5% | 20% |
| Level 3 | 30% | 5% | 7% | 30% |
| Level 4 | 40% | 7% | 10% | 50% |
| Level 5 | 50% | 10% | 15% | 80% |

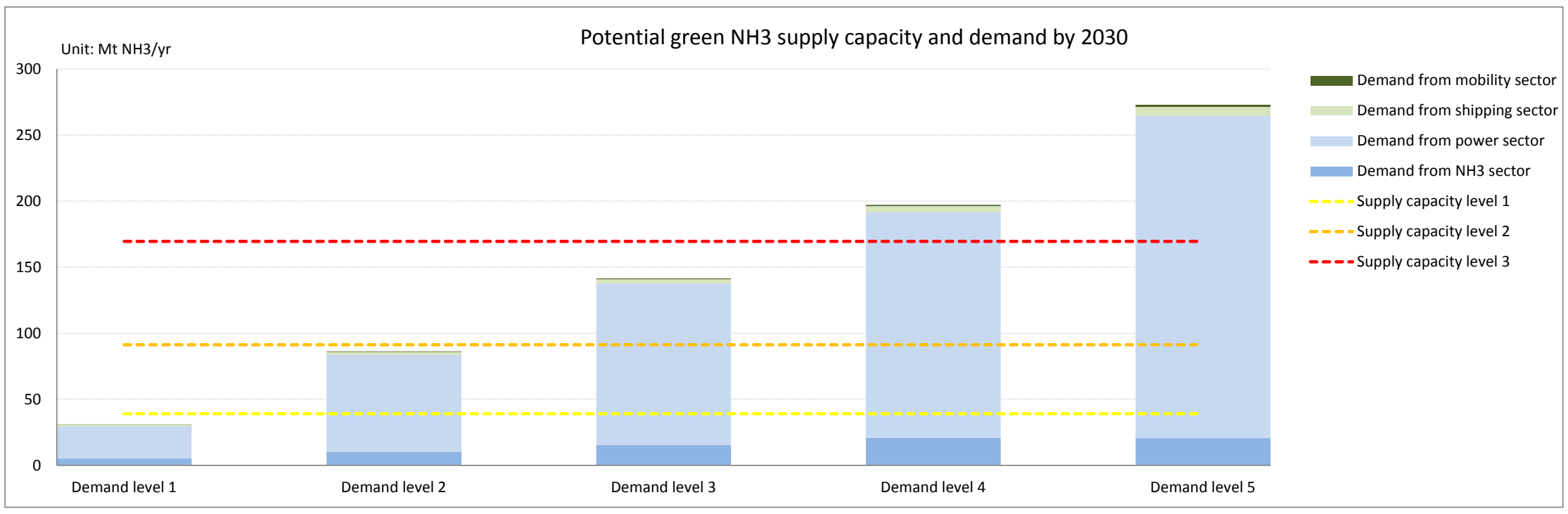

**Fig. 15.** Potential green ammonia supply capacity and demand by 2030

## 4.4 Sustainability of a green ammonia economy

The sustainability of developing the future green ammonia industry in China is analysed from the 3E perspective, by briefing the main issues in China's energy transition, industry decarbonisation and regional development, and discussing the expected advantages of developing a green ammonia industry.

### 4.4.1 Energy

From the energy aspect, the key challenges in China's energy transition are summarized as below:

(1) Coal-dominated energy structure. China planned to slow down the growth of energy consumption and phase out fossil fuel use, however, the primary energy consumption experienced a constant increase in the past decade, and China's economy remains closely tied to fossil energy.
(2) High vulnerability in energy security. Beside coal consumption, demand of crude oil and natural gas gradually increases. Due to the limited domestic supply capacity, more than 70% of oil and 40% natural gas supply depend on imports, which threaten capability of long-term and sustainable energy supply. In addition, the limited energy importers and energy transportation corridors have further worsened China's energy security.
(3) Low renewable energy utilization. China's renewable energy sector grows fast and accounts for 23% of total power generation capacity. However, the divergence between installed capacity and actual power generation grows dramatically due to the spatial mismatch between energy supply and demand. China pursues inter-regional power transmission to facilitate the power transmission to load centres, however, failed to meet its expectation. The intermittent power generation has resulted in low-level transmission and creation of extra thermal power capacity. In addition, the transmission capacity is far from sufficient.

The expected advantages of developing the green ammonia industry from the energy perspective are summarized as follow.

(1) Declining the share of fossil fuels in overall energy consumption. As a potential hydrogen carrier and clean fuel, massive renewable energy can be converted and stored in the form of ammonia which can be further consumed for power generation, heat supply and other purposes. In particular, this can improve the coal-dominated energy structure by lowering the share of coal in the overall primary energy consumption.
(2) Enhancing energy security. Oil and natural gas supply is by far heavily dependent on imports subject to high geopolitical risks. The use of green ammonia as a fuel can reduce the reliance on the imports of oil and natural gas, therefore enhancing the China's energy security.
(3) Enabling large-scale and long-distance hydrogen distribution and long-term hydrogen storage. As ammonia is considered to be a safe and economic medium for hydrogen transportation and storage, and a mature infrastructure already exists, green ammonia can play a key role for large-scale energy distribution and long-term energy storage. As a result, ammonia can contribute to addressing the challenges of intermittent renewable power generation and facilitating inter-regional power transmission and energy distribution. In addition, since ammonia industry has a wide distribution in China, green ammonia is crucial to reduce and tackle the renewable power curtailment especially in the north of China.

### 4.4.2 Environment

From the environmental aspect, the key challenges in China's industry decarbonization are summarized as below:

(1) Massive $CO_2$ emission due to the industrialization. The coal-dominated energy structure has posed a big challenge to China's goals towards the low-carbon energy transition. Power and heat supply emit around half of the overall $CO_2$ in China, and around 70% of which is consumed by the industry sector. In addition, the direct emission from industry accounts for more than 30% of the overall $CO_2$ emission. These indicate that approximately 70% of $CO_2$ emission is caused by China's industrialization, especially after joining the WTO in 2001. Therefore, the industry decarbonisation is essential in achieving China's carbon reduction goals.

(2) The traditional ammonia industry as a representative of one of the most energy-intensive and pollutant industries. The conventional ammonia sector accounts for 40% of global production capacity due to the significant demand mainly from the agricultural and industrial sectors [92]. Ammonia industry is widely distributed in each region, especially in north, east, and mid-south of China. Ammonia is primarily produced from coal and natural gas, especially coal-based ammonia plants is dominant accounting for 76% of production capacity in China. As a result, the ammonia industry is one of the largest energy consumers and most pollutant sectors in China, which has significantly contribute to energy and environmental issues.

The expected advantages of developing the green ammonia industry from the environmental perspective are summarized as follow.

(1) Facilitation of industry decarbonisation by using green $NH_3$ as a clean fuel. As analysed in Section 3.1, around 70% of $CO_2$ emission is caused by industrial production directly and indirectly, and coal is still the dominated energy source in China. The adoption of green ammonia in energy-consuming and pollutant power generation and industrial processes as a clean fuel instead of coal can facilitate industry decarbonisation and energy transition.

(2) Decarbonising the traditional ammonia production. Since conventional ammonia is exclusively produced from fossil energy and the industry is highly energy-intensive and pollutant, the shift to green ammonia can reduce the reliance on fossil fuels (especially on coal), and decarbonize the industry.

### 4.4.3 Economy

From the economic aspect, the main issues of China's regional sustainable development are summarized as follow.

(1) Imbalanced and unsustainable regional development. China's economy has grown as the second largest in the world, but the economic growth is imbalanced, and some regions experience unsustainable development. East and mid-south of China are the most developed regions with high GTFPs and low energy and carbon intensities. In contrast, northwest, southwest, and northeast of China are unsustainable regions with low productivity and high energy consumption and pollutions as per economic size.

The expected advantages of developing the green ammonia industry from the economic perspective are summarized as follow.

(1) Advancing the development of a hydrogen economy. Ammonia is a flexible and economic hydrogen carrier and clean fuel, in addition to being a widely used chemical product. Transporting and storing hydrogen as ammonia can further reduce the cost. Without cracking back to hydrogen, the direct use of ammonia can further reduce the cost and decarbonize the industry sector, and is easy to gain economies of scale effect. Further, the re-use of current ammonia supply chains existing for 10 decades for hydrogen transportation and storage can significantly save the cost in new hydrogen infrastructure investment. Therefore, these are essential to advance the hydrogen economy, especially in the early stage.

(2) Facilitating synergetic regional development. The concept and technology of power-to-x (P2X) is expected to transform and reshape the industrial system, especially phasing out the current coal and oil-chemical industries in China. Currently, around a half of the ammonia production capacity is distributed in the most developed regions, incl. east and mid-south of China. A geographical restructuring of the ammonia industry is inevitable with more production capacity moving close to renewable power bases mainly in north, northwest, northeast and southwest of China. The restructuring would bring a massive transfer of production factors, incl. capital, labour force, and technology, to these regions which can facilitate the local economic development. In addition, these regions would also see a further rise in regional development with the promotion as the supply centre of green ammonia (as well as other green products) in China.

# 5. Conclusions and policy implications

This paper studied the main challenges in China's sustainable development and the role of green ammonia in achieving China's energy and economic development and carbon neutrality. The main findings and results are summarized as follows.

The main challenges in China's energy transition, industry decarbonisation, and sustainable development were explored.

(1) Despite of calling for reducing fossil energy use, the fossil fuel especially coal-dominated energy consumption poses great challenges to achieve energy transition. In addition, the high dependency on oil and gas import with limited importers and energy transportation corridors threatens the energy security. Despite the leading renewable energy investment, the utilization of renewable energy remains low due to spatial energy mismatch and the lack of power market mechanism.

(2) The coal-dominated energy consumption and large-scale industrialization after joining WTO have led to massive CO2 emission. The traditional ammonia industry is one of the largest energy consumers and pollutant industrial sectors, significantly contributing to energy and environmental issues.

(3) Imbalanced and unsustainable regional development were observed by applying indexes of GTFP, carbon and energy intensities in the six administrative regions of China. GTFPs were calculated by applying a DEA optimization model with labour force, capital stock, energy consumption and carbon emission as inputs and GDP as the output. Northwest, northeast and southwest China were evaluated as unsustainable regions with low GTFPs and high energy and carbon intensities, in contrast to East and mid-south China.

To meet the challenges, the role of green ammonia was analysed.

(1) Ammonia was examined to be a flexible and economic option for long-distance and large-volume hydrogen transport and long-term energy storage. Ammonia delivery cost is little influenced by volume and distance (~1.2-1.6 USD/kg H2 for 500-3000 km) and can be further reduced by 45-54% if ammonia is the final feedstock. Storing hydrogen as ammonia is less influenced by storage length (~0.6-0.7 USD/kg H2 for 30-2000 days). In addition, the cost of fuel ammonia was evaluated. Ammonia co-firing in coal power plants at 3-5% rates can be options for China in terms of low-carbon transition, compared to investing into gas power generation. Especially, a 3% co-firing resulting in a 17% increase in levelized cost of electricity is a potential option for the near future. Further, the potential supply capacity and demand of green ammonia by 2030 was evaluated. We found that the conventional ammonia and power sectors can provide sufficient demand. Especially for power generation, a 3% co-firing ammonia in coal power plants can provide significant green ammonia demand (~73 Mt NH3/yr), requiring 28% wind and solar power generation in 2030, due to the large-scale coal power generation.

(2) The sustainability of a green ammonia economy was analysed from energy,

economic and environmental perspectives. From the energy aspect, the use of green ammonia especially as a clean fuel can reduce fossil energy consumption and enhance energy security by lessening the imports of oil and gas. In addition, the use of green ammonia as a hydrogen carrier can facilitate energy transition by enabling large-scale energy distribution and storage. From the environmental aspect, the use of green ammonia as a fertilizer and clean fuel can facilitate industry decarbonisation by phasing out energy-intensive and pollutant production processes and the use of fossil fuels and feedstock. From the economic aspect, the potential of green ammonia being a fertilizer, fuel and energy store can advance the development of a hydrogen economy. In addition, the green ammonia economy can facilitate synergic regional development by driving the industry restructuring with forming new P2X industries and bringing production factors close to renewable resource-rich areas which are mainly located in unsustainably developed regions in China.

In addition, policy implications resulted from this study are presented as follows.

(1) Based on the findings and results, we propose to develop the green ammonia supply chain to meet China's energy, economic and environmental challenges and facilitate the hydrogen economy. First, a green ammonia policy from a national level should be enacted to formulate clear objectives and overall strategies for the industry development, such as the fields of green ammonia utilization, market creation, investment in the industry, etc. Second, regulatory reform is necessary to allow using green ammonia as a hydrogen carrier and a clean fuel. Third, mid and long-term development plans should be developed for developing the industry in different stages. The plan should be in line with the overall policy. Fourth, detailed policies, rules, and regulations for market and infrastructure development should be developed.

(2) The use of fuel ammonia in power generation and transportation should be paid attention as important applications to facilitate the industry development in the early stage. Based on the analysis of this study, the power sector particularly can provide significant demand for green ammonia. China considers increasing the share of natural gas in thermal power generation instead of coal to achieve carbon reduction, since natural gas is a cleaner fossil energy with 68% carbon emission that of coal [93]. In contrast, co-firing ammonia can reuse the existing coal power plants to avoid massive investments in natural gas power plants, meanwhile reducing carbon emission. Further, since natural gas consumption heavily relies on import, co-firing ammonia can secure energy safety by producing ammonia with renewable power, meanwhile addressing renewable power curtailment in China.

Besides, our study has some limitations. First, the levelized hydrogen delivery and storage costs were estimated based on a constant cost and energy delivered or stored per year. The mixed fuel cost and levelized cost of electricity for ammonia co-firing are evaluated based on a set of parameters with average numbers. Therefore, these can vary in the light of actual operating and regional conditions. Second, ammonia co-firing was analysed in terms of supply and demand and cost increase in power generation, however, the impact on demand side, such as the industry and residential sectors has not been considered. In addition, impact of co-firing on fuel and electricity costs does not consider carbon tax. These can be studied to further inform policy.

# Acknowledgements

The author would like to thank China Scholarship Council for the funding provided to carry out this research.

# Nomenclature

| Abbreviation | Full name |
|---|---|
| 3E | Energy-environment-economy |
| Btce | Billon ton of standard coal equivalent |
| CRS | Constant return to scale |
| DEA | Data Envelope Analysis |
| DMU | Decision-making unit |
| GC | Coal gasfication |
| GDP | Gross domestic product |
| GTFP | Green total factor productivity |
| P2G | Power to gas |
| Mtoe | Million ton of standard oil equivalent |
| SMR | Steam reforming |
| WTO | World trade organization |

| Index | Definition |
|---|---|
| $n$ | Year |
| $i$ | DMU |
| $j$ | DMU |

| Variable/Parameter | Definition | Unit |
|---|---|---|
| $\delta$ | Depreciation rate | ea. |
| $\vartheta_i$ | Efficiency of the DMU *i* ranging from 0-1 | ea. |
| $\lambda_i$ | Weight of the DMU *j* | ea. |
| $Capex_n$ | Investment cost in year *n*. | USD/yr |
| $C_{i,n}$ | CO2 emission of region *i* in year *n* | B kg/yr |
| $CI_{i,n}$ | Carbon intensity of region *i* in year *n* | kg/USD |
| $dr$ | Discount rate | % |
| $E_n$ | Quantity of energy delivered or stored in year *n*. | kg/yr |
| $E_{i,n}$ | Energy consumption of region *i* in year *n* | B Btu/yr |
| $EI_{i,n}$ | Energy intensity of region *i* in year *n* | Btu/USD |
| $EP^c$ | Share of coal cost in power generation | % |
| $Exp_n$ | Total expense in year *n* | USD/yr |
| $FC^m$ | Consumption of the mixed fuel for power generation per unit | tce/MWh |
| $FE^{am}$ | Fuel cost for green ammonia | USD/tce |
| $FE^c$ | Fuel cost for thermal coal | USD/tce |
| $FE^m$ | Mixed fuel cost | USD/tce |
| $FR^m$ | Co-firing rate | % |
| $G_{i,n}$ | GDP of region *i* in year *n* | B USD/yr |
| $I_{i,n}$ | Fixed capital stock of year *n* in region *i* | B USD |
| $K_{i,n}$ | Capital stock of year *n* in region *i* | B USD |
| $LCOE^c$ | Levelized cost of coal power | USD/MWh |
| $LCOE^m$ | Levelized cost of electricity in co-firing ammonia with coal for power generation | USD/MWh |
| $Opex_n$ | Operating cost in year *n*. | USD/yr |
| $PR^m$ | Penetration rate of green ammonia in the mobility sector | % |
| $TR^{am}$ | Average utilization rate of hydrogen refilling stations | % |
| $UR^{hrs}$ | The share of hydrogen transported as ammonia | % |
| $X_j$ | the inputs of the DMU *j* | n.a |
| $Y_j$ | the outputs of the DMU *j* | n.a |

# References

[1] Hoggett R, Bolton R, Candelise C, Kern F, Mitchell C, Yan J. Supply chains and energy security in a low carbon transition. Applied Energy. 2014;123:292-5.
[2] Ye Q, Lu J, Zhu M. Wind curtailment in China and lessons from the United States. China's Energy in Transition. 2018.
[3] Li Y, Lukszo Z, Weijnen M. The impact of inter-regional transmission grid expansion on China’s power sector decarbonization. Applied Energy. 2016;183:853-73; https://doi.org/10.1016/j.apenergy.2016.09.006.
[4] Mengye Z, Ye Q, David B, Jiaqi L, Bart K. The China wind paradox: The role of state-owned enterprises in wind power investment versus wind curtailment. Energy Policy. 2019;127:200-12; https://doi.org/10.1016/j.enpol.2018.10.059.
[5] Russell C. Uncertainty is key in IEA's hopes for China's energy transition. Reuters. 2023; https://www.reuters.com/business/energy/uncertainty-is-key-ieas-hopes-chinas-energy-transition-russell-2023-10-24/.
[6] NDRC. Mid and Long-term Hydrogen Development Plan of China (2021-2035). China: National Development and Reform Commission; 2022.
[7] IEA. The future of hydrogen. IEA. 2019; https://www.iea.org/reports/the-future-of-hydrogen.
[8] Li G, Ma Z, Zhao J, Zhou J, Peng S, Li Y, et al. Research progress in green synthesis of ammonia as hydrogen-storage carrier under ‘hydrogen 2.0 economy’. Clean Energy. 2023;7:116-31; https://doi.org/10.1093/ce/zkac095.
[9] Asna Ashari P, Oh H, Koch C. Pathways to the hydrogen economy: A multidimensional analysis of the technological innovation systems of Germany and South Korea. International Journal of Hydrogen Energy. 2024;49:405-21; https://doi.org/10.1016/j.ijhydene.2023.08.286.
[10] Harichandan S, Kar SK, Rai PK. A systematic and critical review of green hydrogen economy in India. International Journal of Hydrogen Energy. 2023;48:31425-42; https://doi.org/10.1016/j.ijhydene.2023.04.316.
[11] Chu KH, Lim J, Mang JS, Hwang M-H. Evaluation of strategic directions for supply and demand of green hydrogen in South Korea. International Journal of Hydrogen Energy. 2022;47:1409-24; https://doi.org/10.1016/j.ijhydene.2021.10.107.
[12] Li C, Zhang L, Ou Z, Ma J. Using system dynamics to evaluate the impact of subsidy policies on green hydrogen industry in China. Energy Policy. 2022;165:112981; https://doi.org/10.1016/j.enpol.2022.112981.
[13] Choi J, Choi DG, Park SY. Analysis of effects of the hydrogen supply chain on the Korean energy system. International Journal of Hydrogen Energy. 2022;47:21908-22; https://doi.org/10.1016/j.ijhydene.2022.05.033.
[14] Jia D, Li X, Gong X, Lv X, Shen Z. Bi-level strategic bidding model of novel virtual power plant aggregating waste gasification in integrated electricity and hydrogen markets. Applied Energy. 2024;357:122468; https://doi.org/10.1016/j.apenergy.2023.122468.
[15] Schlund D, Schulte S, Sprenger T. The who’s who of a hydrogen market ramp-up: A stakeholder analysis for Germany. Renewable and Sustainable Energy Reviews.

2022;154:111810; https://doi.org/10.1016/j.rser.2021.111810.
[16] Khan MI, Al-Ghamdi SG. Hydrogen economy for sustainable development in GCC countries: A SWOT analysis considering current situation, challenges, and prospects. International Journal of Hydrogen Energy. 2023;48:10315-44; https://doi.org/10.1016/j.ijhydene.2022.12.033.
[17] Huang Y, Zhou Y, Zhong R, Wei C, Zhu B. Hydrogen energy development in China: Potential assessment and policy implications. International Journal of Hydrogen Energy. 2024;49:659-69; https://doi.org/10.1016/j.ijhydene.2023.10.176.
[18] Hong S, Kim E, Jeong S. Evaluating the sustainability of the hydrogen economy using multi-criteria decision-making analysis in Korea. Renewable Energy. 2023;204:485-92; https://doi.org/10.1016/j.renene.2023.01.037.
[19] Lee Y, Cho MH, Lee MC, Kim YJ. Policy agenda toward a hydrogen economy: Institutional and technological perspectives. International Journal of Hydrogen Energy. 2024;54:1521-31; https://doi.org/10.1016/j.ijhydene.2023.12.129.
[20] Palacios A, Palacios-Rosas E, De alburqueque R, Castro-Olivera PM, Oros A, Lizcano F, et al. Hydrogen in Mexico: A technical and economic feasibility perspective for the transition to a hydrogen economy. International Journal of Hydrogen Energy. 2024; https://doi.org/10.1016/j.ijhydene.2024.03.116.
[21] Gujarathi AM, Al-Hajri R, Al-Ani Z, Al-Abri M, Al-Rawahi N. Towards technology, economy, energy and environment oriented simultaneous optimization of ammonia production process: Further analysis of green process. Heliyon. 2023;9:e21802; https://doi.org/10.1016/j.heliyon.2023.e21802.
[22] Wu Y, Zhao T, Tang S, Wang Y, Ma M. Research on design and multi-frequency scheduling optimization method for flexible green ammonia system. Energy Conversion and Management. 2024;300:117976; https://doi.org/10.1016/j.enconman.2023.117976.
[23] Zhao H, Kamp LM, Lukszo Z. Exploring supply chain design and expansion planning of China's green ammonia production with an optimization-based simulation approach. International Journal of Hydrogen Energy. 2021;46:32331-49; https://doi.org/10.1016/j.ijhydene.2021.07.080.
[24] Zhao H, Kamp LM, Lukszo Z. The potential of green ammonia production to reduce renewable power curtailment and encourage the energy transition in China. International Journal of Hydrogen Energy. 2022;47:18935-54; https://doi.org/10.1016/j.ijhydene.2022.04.088.
[25] Tu R, Liu C, Shao Q, Liao Q, Qiu R, Liang Y. Pipeline sharing: Optimal design of downstream green ammonia supply systems integrating with multi-product pipelines. Renewable Energy. 2024;223:120024; https://doi.org/10.1016/j.renene.2024.120024.
[26] Zhao H. Green ammonia supply chain and associated market structure. Fuel. 2024;366:131216; https://doi.org/10.1016/j.fuel.2024.131216.
[27] Joseph Sekhar S, Samuel MS, Glivin G, Le TG, Mathimani T. Production and utilization of green ammonia for decarbonizing the energy sector with a discrete focus on Sustainable Development Goals and environmental impact and technical hurdles. Fuel. 2024;360:130626; https://doi.org/10.1016/j.fuel.2023.130626.
[28] Morlanés N, Katikaneni SP, Paglieri SN, Harale A, Solami B, Sarathy SM, et al. A technological roadmap to the ammonia energy economy: Current state and missing

technologies. Chemical Engineering Journal. 2021;408:127310; https://doi.org/10.1016/j.cej.2020.127310.
[29] Galimova T, Fasihi M, Bogdanov D, Breyer C. Feasibility of green ammonia trading via pipelines and shipping: Cases of Europe, North Africa, and South America. Journal of Cleaner Production. 2023;427:139212; https://doi.org/10.1016/j.jclepro.2023.139212.
[30] Miller CA, Iles A, Jones CF. The Social Dimensions of Energy Transitions. Science as Culture. 2013;22:135-48; https://doi.org/10.1080/09505431.2013.786989.
[31] Hartwig M, Emori S, Asayama S. Normalized injustices in the national energy discourse: A critical analysis of the energy policy framework in Japan through the three tenets of energy justice. Energy Policy. 2023;174:113431; https://doi.org/10.1016/j.enpol.2023.113431.
[32] Jefferson M. A global energy assessment. WIREs Energy and Environment. 2016;5:7-15; https://doi.org/10.1002/wene.179.
[33] CNKI. National economic and social development statistics. CNKI. 2020; https://data.cnki.net/.
[34] BP. Statistical Review of World Energy 2021. 2021.
[35] Liu SQ, Huang X, Li X, Masoud M, Chung S-H, Yin Y. How is China’s energy security affected by exogenous shocks? Evidence of China–US trade dispute and COVID-19 pandemic. Discover Energy. 2021;1:2; 10.1007/s43937-021-00002-6.
[36] Energy security faces three challenges. Energy Law. 2019; http://energylaw.chinalaw.org.cn/portal/article/index/id/2248.html.
[37] Ye Q. China’s peaking emissions and the future of global climate policy. China's Energy in Transition. 2018.
[38] IRENA. Renewable Capacity Statistics 2021. International Renewable Energy Agency; 2021.
[39] Downie E. Sparks fly over ultra-high voltage power lines 2018.
[40] Xi J. Wind power transmission problem in the three North regions of China remains unsolved. 2020; https://www.jiemian.com/article/5263167.html.
[41] Shan Y, Guan D, Zheng H, Ou J, Li Y, Meng J, et al. China CO2 emission accounts 1997–2015. Scientific Data. 2018;5:170201; 10.1038/sdata.2017.201.
[42] CEADs. China's carbon emission from 2014-2019. Carbon emission accounts & datasets. 2021; https://www.ceads.net.cn/news/20211256.html.
[43] NEA. China's power consumption in 2019. National Energy Administration. 2020; http://www.nea.gov.cn/2020-01/20/c_138720881.htm.
[44] CHIC. Industry report 2020. clean heat supply: Clean Heating Industry Commitee; 2020.
[45] CEIC. Number of Registered Vehicles in China. CEIC Data. 2023; https://www.ceicdata.com/en/indicator/china/number-of-registered-vehicles.
[46] CEC. Economic development of the power industry in 2019. China Electricity Council; 2020.
[47] CHIC. The development of clean heating industry. 2021; https://www.chic.org.cn/home/Index/detail1?id=1125.
[48] Verheul B. Overview of hydrogen and fuel cell developments in China. Shanghai, China: Holland Innovation Network; 2019.
[49] Why China wants to develop hydrogen energy. CEBM. 2016; http://www.escn.co

m.cn/news/show-373397.html.
[50] Ma D, Hasanbeigi A, Chen W. Energy-efficiency and air-pollutant emissions-reduction opportunities for the ammonia industry in china. Lawrence Berkeley National Laboratory.
[51] Wen Q. Overview of China's ammonia industry in the past 40 years. China's National Petroleum & Chemical Planning Institute. 2018; http://www.ciccc.com/Content/2019/01-02/0958091216.html.
[52] D'Adamo I, Gastaldi M, Imbriani C, Morone P. Assessing regional performance for the Sustainable Development Goals in Italy. Scientific Reports. 2021;11:24117; https://doi.org/10.1038/s41598-021-03635-8.
[53] Yu B, Fang D, Pan Y, Jia Y. Countries' green total-factor productivity towards a low-carbon world: The role of energy trilemma. Energy. 2023;278:127894; https://doi.org/10.1016/j.energy.2023.127894.
[54] Cui L, Mu Y, Shen Z, Wang W. Energy transition, trade and green productivity in advanced economies. Journal of Cleaner Production. 2022;361:132288; https://doi.org/10.1016/j.jclepro.2022.132288.
[55] Zhang J, Wu G, Sun J. The estimation of China's provincial capital stock: 1952-2000. Economic Research Journal. 2004;33.
[56] Zhong Z, Liao Z. Estimates of fixed capital stock by sector and region: 1978-2011. Journal of Guizhou University of Finance and Economics. 2014;32:8-9.
[57] CEADs. Carbon emission in Tibet and associated cities in 2014. Carbon emission accounts & datasets. 2021; https://www.ceads.net.cn/data/province/by_sectoral_accounting/tibet2014/.
[58] NDRC. High-quality development in the condition of achieving carbon reduction goals. National Development and Reform Commission. 2021; https://www.ndrc.gov.cn/wsdwhfz/202112/t20211231_1311184.html.
[59] Qingyi W. Energy data 2020. Institute for Global Decarbonization Progress; 2020.
[60] Liu S, Bie Z, Lin J, Wang X. Curtailment of renewable energy in Northwest China and market-based solutions. Energy Policy. 2018;123:494-502; https://doi.org/10.1016/j.enpol.2018.09.007.
[61] Cesaro Z, Ives M, Nayak-Luke R, Mason M, Bañares-Alcántara R. Ammonia to power: Forecasting the levelized cost of electricity from green ammonia in large-scale power plants. Applied Energy. 2021;282:116009; https://doi.org/10.1016/j.apenergy.2020.116009.
[62] Bartels JR, Pate MB. A feasibility study of implementing an Ammonia Economy. 2008.
[63] Aolin H, Chunlei H, Yufeng S, Zhibo Y. A discussion on the pricing mechanism of underground gas storage in China. Natural Gas Industry. 2010;30:91-6.
[64] Thomas D. State of play and developments of power-to-hydrogen technologies Hydrogenics Europe; 2019.
[65] Alibaba. Alibaba sales platform. Alibaba. 2021; https://www.alibaba.com/.
[66] Elizabeth Connelly, Michael Penev, Amgad Elgowainy, Hunter C. Current Status of Hydrogen Liquefaction Costs. DOE; 2019.
[67] Si W, Yu D. How much will the cost of hydrogen be decreased in China? Hydrogen transport and storage. Guangzhou, China: GuangZheng HengSheng; 2019.
[68] IEA. Ammonia technology roadmap. IEA; 2021.
[69] DOE. Renewable energy to fuels through utilization of energy-dense liquids (refuel). U.S.:

U.S. Department of Energy; 2016.
[70] Connor N. Hydrogen - latent heat of vaporization. Nuclear Power. 2021; https://www.nuclear-power.com/hydrogen-specific-heat-latent-heat-vaporization-fusion/.
[71] Al-Breiki M, Bicer Y. Technical assessment of liquefied natural gas, ammonia and methanol for overseas energy transport based on energy and exergy analyses. International Journal of Hydrogen Energy. 2020;45:34927-37; https://doi.org/10.1016/j.ijhydene.2020.04.181.
[72] TrendBank. Liquid hydrogen industry and its future trends. TrendBank. 2021; https://news.bjx.com.cn/html/20210621/1159193.shtml.
[73] Bellini E. Green hydrogen and the cable-pipeline dilemma. PV Magazine. 2021; https://www.pv-magazine.com/2021/04/19/green-hydrogen-and-the-cable-pipeline-dilemma/.
[74] Liu Z, Zhao Y, Pan L. Analysis and comparison of energy saving efficiency and emission reduction efficiency of thermal power between China and foreign countries. Thermal Power Generation. 2021;50:9-18; https://doi.org/10.19666/j.rlfd.202009232.
[75] China market price of liquid ammonia. CEIC. 2021; https://www.ceicdata.com/en/china/china-petroleum--chemical-industry-association-petrochemical-price-fertilizer/cn-market-price-monthly-avg-fertilizer-liquid-ammonia-998-or-above.
[76] Burgess J, Garg V. Asia-Pacific blue ammonia prices extend discount to Europe, US. S&P Global. 2024; https://www.spglobal.com/commodityinsights/en/market-insights/latest-news/energy-transition/022824-asia-pacific-blue-ammonia-prices-extend-discount-to-europe-us.
[77] RootsAnalysis. Green Ammonia Market. RootsAnalysis. 2024; https://www.rootsanalysis.com/reports/green-ammonia-market.html.
[78] FutureBridge. Green ammonia - an alternative fuel. FutureBridge. 2022; https://www.futurebridge.com/industry/perspectives-energy/green-ammonia-an-alternative-fuel/.
[79] BP. Statistical Review of World Energy 2022. BP. 2022; https://www.bp.com/content/dam/bp/business-sites/en/global/corporate/pdfs/energy-economics/statistical-review/bp-stats-review-2022-full-report.pdf.
[80] Dias V, Pochet M, Contino F, H J. Energy and economic costs of chemical storage. Frontiers in Mechnical Engineering. 2020;6; https://10.3389/fmech.2020.00021.
[81] Caculation of coal power generation cost. BJX. https://news.bjx.com.cn/html/20170522/826657-2.shtml.
[82] Xu Y, Wang H, Liu X, Zhu J, Xu J, Xu M. Mitigating CO2 emission in pulverized coal-fired power plant via co-firing ammonia: A simulation study of flue gas streams and exergy efficiency. Energy Conversion and Management. 2022;256:115328; https://doi.org/10.1016/j.enconman.2022.115328.
[83] Zhang Q. Grid parity of renewable electricity. BJX. 2019; https://guangfu.bjx.com.cn/news/20190328/971678-3.shtml.
[84] SCC. China Economic Census Yearbook. The State Council of China; 2018.
[85] Zhao F, Mu Z, Hao H, Liu Z, He X, Victor Przesmitzki S, et al. Hydrogen Fuel Cell Vehicle Development in China: An Industry Chain Perspective. Energy Technology. 2020;8:2000179; https://doi.org/10.1002/ente.202000179.
[86] Paul P. Renewable Hydrogen Transportation Fuel Production. California Energy

Commission; 2020.
[87] Lhrating. Thermal power industry in China and outlook. China Lianhe Credit Rating; 2021.
[88] Brown T. Japan's Road Map for Fuel Ammonia. Ammonia Energy. 2021; https://www.ammoniaenergy.org/articles/japans-road-map-for-fuel-ammonia/.
[89] Zhang Y. China's hydrogen industrial report. China EV100; 2020.
[90] SGERI. China Energy and Power Outlook. State Grid Energy Reserach Institute; 2020.
[91] Wang W. Power system and renewable energy in China. Guosen Securities; 2022.
[92] Zeng C. Overview of China ammonia industry. Ammonia Plant Safety and Related Facilities. 2014;55:119-25.
[93] GE. Natural gas power generation to promote low-carbon transition and enter the hydrogen era. GE; 2021.